\documentclass[a4paper,12pt]{article}

\usepackage{jheppub}

\usepackage{amssymb}
\usepackage{graphicx}
\usepackage{amsmath,amsfonts}
\usepackage{,color}
\usepackage{epsfig}
\usepackage{bbm}
\usepackage{hyperref}

\def\l{\left(}
\def\r{\right)}
\def\la{\langle }
\def\ra{\rangle }
\newcommand{\be}{\begin{equation}}
\newcommand{\ee}{\end{equation}}
\newcommand{\bea}{\begin{eqnarray}}
\newcommand{\eea}{\end{eqnarray}}
\newcommand{\bg}{\begin{gather}}
\newcommand{\eg}{\end{gather}}
\newcommand{\bseq}{\begin{subequations}}
\newcommand{\eseq}{\end{subequations}}

\renewcommand{\ln}{\mathop{\rm ln}\nolimits}

\def\half{\frac{1}{2}}

\newcommand{\bra}[1]{\langle #1 |}
\newcommand{\ket}[1]{| #1 \rangle}

\newcommand{\cbh}{\ensuremath{C_\text{BH}} }

\DeclareMathOperator{\tr}{tr}

\newcommand{\eps}{\ensuremath{\varepsilon}}

\usepackage{soul}

\title{Density matrix of black hole radiation}
\author{Lasma Alberte,$^{\textnormal{a}}$ }
\author{Ram Brustein,$^{\textnormal{a}}$ }
\author{Andrei Khmelnitsky,$^{\textnormal{a}}$ }
\author{and A.J.M. Medved$^{\textnormal{b,c}}$}
\affiliation{$^{\textrm{a}}$ Department of Physics, Ben-Gurion University, Beer Sheva 84105, Israel}
\affiliation{$^{\textrm{b}}$ Department of Physics $\&$ Electronics, Rhodes University, Grahamstown 6140, South Africa}
\affiliation{$^{\textrm{c}}$ National Institute for Theoretical Physics (NITheP), Western Cape 7602, South Africa}
\emailAdd{lasma@post.bgu.ac.il}
\emailAdd{ramyb@bgu.ac.il}
\emailAdd{andreykh@post.bgu.ac.il}
\emailAdd{j.medved@ru.ac.za}

\abstract{Hawking's model of black hole evaporation is not unitary and leads to a mixed density matrix for the emitted radiation, while the Page model describes a unitary evaporation process in which the density matrix evolves from an almost thermal state to a pure state. We compare a recently proposed model of semiclassical black hole evaporation to the two established models. In particular, we study the density matrix of the outgoing radiation and determine how the magnitude of the off-diagonal corrections differs for the three frameworks. For Hawking's model, we find   power-law corrections to the two-point functions that induce exponentially suppressed corrections to the off-diagonal elements of the full density matrix. This verifies that the Hawking result is correct to all orders in perturbation theory and  also allows one to express the full  density matrix in terms of the single-particle density matrix. We then consider the semiclassical theory for which the corrections, being  non-perturbative from an  effective field-theory perspective, are much less suppressed and grow monotonically in time. In this case, the R\'enyi entropy for the outgoing radiation is shown to grow linearly at early times; but this growth slows down and the entropy eventually starts to decrease at the Page time. In addition to comparing models, we emphasize the distinction between the state of the radiation emitted from a black hole, which is highly quantum, and that of the radiation emitted from a typical classical black body at the same temperature.}

\begin{document}

\maketitle
\flushbottom

\section{Introduction}

There has been a recent spike in activity on understanding the implications of black hole (BH) evaporation~\cite{Hawking:1974sw,Hawking:1976ra}. This can be attributed, in large part, to the controversial proposal that a unitary evaporation process comes with the cost of a ``firewall''~\cite{AMPS} or ``energetic curtain''~\cite{Braun}---these being colorful euphemisms for an apparent tension between general relativity and the unitarity of quantum field theory in a BH setting~\cite{Sunny,Mathur,MP,Bousso}. While the debate rages on, a consensual mechanism for information release is still lacking.

In Hawking's model, the process of BH evaporation is not unitary~\cite{Hawking:1976ra}. Hawking argued that the correlation functions of the emitted radiation are diagonal in mode-occupancy number, frequency and emission time and, from
this,  deduced that the density matrix for the radiation is diagonal in the  same quantities. In fact, to  good accuracy, the final state of the emitted radiation is thermal, and so it is similar to a maximally mixed state. Then the evaporation cannot be the result of unitary evolution from a nearly pure state, which lead Hawking to state (a statement which he more recently retracted~\cite{retraction}) that the process of gravitational collapse is not compatible with the standard principles of quantum mechanics.

The established benchmark model for describing unitary BH evaporation has been the Page model~\cite{Page}. In this model, the BH and the emitted radiation are assumed to be in a pure state in some large Hilbert space, which is partitioned into an ``in" part representing the BH and an ``out" part representing the outgoing  radiation. Because the combined state is pure, there must be a special basis in which the full density matrix has only a single eigenvalue. An
external observer measures the reduced ``out" density matrix in some random basis and treats the transformation matrix $U$ as a random matrix with prescribed statistics~\cite{Page,Lubkin}. (See also~\cite{Harlow:2014yka} for a recent review.) When the ``out" system is smaller than the ``in" system, it looks to an
external observer as if it were thermal. Conversely,  when the ``out" system becomes
the  larger one, the deviations from a thermal state grow and indicate that the total state is indeed pure. The critical time in which the midpoint of
evaporation is reached is normally called the Page time. Given that the BH is initially in a pure state  and
that the  evolution is unitary, the
 Page model can plausibly be viewed as setting the minimal rate for purification of the emitted radiation.

Two of the current authors (RB and AJMM) have recently proposed a semiclassical model of BH evaporation~\cite{Brustein:2013qma,Brustein:2013ena}, with the premise of repeating Hawking's seminal calculations~\cite{Hawking:1974sw,Hawking:1976ra}---which assume a classical background metric---so as to include a fluctuating BH geometry. The incipient BH is endowed with a quantum wavefunction~\cite{RM,RB}, leading to expectation values in place of fixed classical parameters. From this perspective, Hawking's model pertains to the limit of an infinitely massive BH with a fixed size, whereas the semiclassical reformulation treats the BH mass as finite with a continually decreasing size due to classical back-reaction effects. In the new model, the correlation functions of the emitted radiation are no longer diagonal and the evaporation process becomes unitary, even though
the thermal-like emission spectrum is maintained. This model of unitary BH evaporation obeys the general constraints about the rate of purification. In particular, it has been shown in~\cite{endgame},  relying on the results of the current paper, that the R\'enyi entropy of the radiation decreases monotonically from its peak value at the Page time  and does so at a faster rate than the Page model predicts.

From the point of view of an  effective field theory in a fixed curved-space background, the correlation functions of the semiclassical model are modified in a non-perturbative way~\cite{RB}. Contrary to expectations, these modifications could become significant in some situations. It has been shown that, when the leading result in the effective field theory either vanishes or diverges,  then the non-perturbative corrections are particularly relevant~\cite{fluckyou}.
In the first case, we expect a correction of the order $1/S_{BH}$ (the inverse
of the BH entropy) to become the leading  result and, in the second, the divergence should be replaced by a large but finite $S_{BH}$. These small non-perturbative quantum effects could be coherent, and so their amplitudes could add up and grow with time;
even growing to the point where they become comparable to the classical outcome
\cite{Brustein:2013qma}.

The previous studies of the new model have utilized the single-particle density matrix (the two-point function) of the radiation as the primary tool\footnote{This matrix was called ``the density matrix" by abuse of language or ``the radiation matrix" in the previous studies.} rather than the full multi-particle density matrix. The single-particle density matrix is easier to calculate and, for the radiation emitted by a BH,  sufficient to completely determine the full density matrix. One of the primary goals of the current paper is to make explicit the relationship between the single-particle and multi-particle density matrices. The results are presented in a way that can be applied to the Hawking
model, the proposed semiclassical model and even to more general modifications of   Hawking's framework. As an upshot,  we are able to find a closed expression for the R\'enyi entropy of the full  density matrix that is expressed directly in terms of the single-particle density matrix.

It has been argued  for BHs in Anti-de Sitter (AdS) space---at first by Maldacena~\cite{eternal} with subsequent elaborations by  many others --- that non-perturbative contributions from other geometries which do not possess a horizon could be relevant to the state of the Hawking radiation. Maldacena's specific example was the contribution of thermal AdS to the matter correlations for large BHs. In this case, the contribution to the density matrix of the radiation is expected to be exponentially suppressed $\sim e^{-S_{BH}}$. However, because the number of off-diagonal elements is exponentially large $\sim e^{+S_{BH}}$, one could not be sure as to the relevance of such a contribution. It could  then be argued that, without a better control on the non-perturbative contributions from other geometries (which are usually not available), it becomes difficult to trust the Hawking result.

We have two comments on this issue. First, according to our analysis, one requires much larger corrections $\sim e^{-S_{BH}/2}$ to challenge the reliability of the leading-order outcomes (See, for example,~\cite{eliezer}). Second, the quantum fluctuations of the background BH geometry induce off-diagonal effects in frequency space that are only power suppressed $\sim 1/\sqrt{S_{BH}}$ and sufficient by themselves to restore unitarity. These statements will be made more precise in due course.

The rest of the paper is organized as follows: In the next section, we review the basic framework of the Hawking model of BH evaporation and recall its main results. Then, in Section~\ref{sec:denmat}, we formulate a precise relationship between the multi-particle density matrix and the single-particle density matrix. The preceding analysis  is applied to the Hawking model in
Section~\ref{sec:state}, with the aim of regularizing  formal divergences
that arise in the eternal-BH framework.
Section~\ref{sec:state} also includes a discussion on how the state of the BH radiation differs from that of thermal radiation emitted by a ``normal" black body. The formal analysis up to this point sets the stage for Section~\ref{sec:off_diagonal}, where we discuss similarities and differences between the three models of BH evaporation: the Hawking model, the Page model and the semiclassical model. There is then a brief conclusion in Section~\ref{sec:conclusion}, followed by four appendices that fill in some technical gaps in the main text.

\section{The Hawking model of black hole evaporation}

Let us start here by recalling the original Hawking description of BH evaporation, which completely dismisses the back-reaction and time-dependence effects. This paper uses the notations of Hawking in~\cite{Hawking:1976ra}. In particular, we consider the initial vacuum state $\ket{0_-}$ and a final vacuum state $\ket{0_+}$. The initial vacuum contains no particles with respect to the creation and annihilation operators $\{a_i^+,\,a_i\}$ as defined at past null infinity,
\be
a_i\ket{0_-}=0\;.
\ee

Following Hawking, the Hilbert space at future infinity is a direct product of the Hilbert space of the particles falling into the BH, $\mathcal H_{\textrm{in}}$, and of those going out, $\mathcal H_{\textrm{out}}$, so that
$\mathcal H=\mathcal H_{\textrm{in}}\otimes\mathcal H_{\textrm{out}}$.
The ingoing particles are created and annihilated by the operators $\{c_i^+,\,c_i\}$, whereas the outgoing ones are created and annihilated by $\{b_i^+,\,b_i\}$. Both are related to $\{a_i^+,\,a_i\}$ via Bogolyubov transformations. The final vacuum state can be represented as $\ket{0_+}=\ket{0_{\textrm{in}}}\ket{0_{\textrm{out}}}$ and contains neither outgoing nor ingoing particles,
\be
b_i\ket{0_+}=0\;,\qquad c_i\ket{0_+}=0\;.
\ee

The two vacuum states $\ket{0_-}$ and $\ket{0_+}$ do not coincide with one another due to the particle creation by the BH. In other words, an initially empty state will appear to have a non-zero occupation number in the final state, meaning that $b_{i}\ket{0_-}\neq 0$. The initial vacuum can be expressed as a linear combination of the Fock states at future infinity as
\be
\ket{0_-}=\lambda_{ab}\ket{a_{\rm out}}\ket{b_{\rm in}}\;,
\ee
where $\ket{a_{\rm out}}$ and $\ket{b_{\rm in}}$ are the basis vectors of the Fock spaces which are spanned by the operators $\{b_i^+,\,b_i\}$ and $\{c_i^+,\,c_i\}$, respectively. As a consequence,
\begin{align}
&\ket{a_{\textrm{out}}}=\prod_j\frac{1}{\sqrt{n_{ja}!}}\left(b^+_j\right)^{n_{ja}}\ket{0_\textrm{out}}\;,\,
&\ket{b_{\textrm{in}}}=\prod_k\frac{1}{\sqrt{n_{kb}!}}\left(c^+_k\right)^{n_{kb}}\ket{0_\textrm{in}}\;.
\end{align}

The object of interest is the density matrix of the initial vacuum state,
\be
\widehat\rho_{\rm vac}\equiv\ket{0_-}\bra{0_-}\;,
\ee
which allows one to calculate the vacuum expectation values of observables according to the standard rule $\bra{0_-}\widehat{\mathcal O}\ket{0_-}=\mathrm {tr}\,\widehat\rho_{\rm vac}\,\widehat{\mathcal O}$.

In particular, we will be interested in the expectation values of the observables at future infinity. These will be composed only from the outgoing creation and annihilation operators $\{b_i^+,\,b_i\}$ and so can be written as $\widehat{\mathcal O}=\widehat{\mathcal O}_{\textrm{out}}\otimes\widehat{\mathcal I}_{\textrm{in}}$, where $\widehat{\mathcal I}_{\textrm{in}}$ denotes the identity operator in the Hilbert space of the ingoing states. The expectation value of this operator then takes the form
\be
\bra{0_-}\widehat{\mathcal O}\ket{0_-}=\bra{0_-}\widehat{\mathcal O}_{\textrm{out}}\ket{0_-}={\rho^{\textrm{\; out}}}_{ac}\,{\mathcal O_{\textrm{out}}}^{ca}=\textrm{tr}_{\textrm{out}}\widehat\rho^{\textrm{\; out}} \widehat{\mathcal O}_{\textrm{out}}\;.
\ee
Here, ${\rho^{\textrm{out}}}_{ac}$ are the matrix elements of the reduced density matrix of the Hilbert space $\mathcal H_{\textrm{out}}$  for the outgoing radiation and are determined by
\be\label{dm}
\widehat\rho^{\textrm{\; out}}\equiv\textrm{tr}_{\textrm{in}}\widehat\rho_{\rm vac}= \bra{b_{\textrm{in}}}0_-\rangle \langle 0_-\ket{b_{\textrm{in}}}=\lambda_{ab}\bar\lambda_{cb} \ket{a_{\textrm{out}}}\bra{c_{\textrm{out}}}={\rho^{\textrm{out}}}_{ac} \ket{a_{\textrm{out}}}\bra{c_{\textrm{out}}}\;,
\ee
whereas
\be
{\mathcal O_{\textrm{out}}}^{ca}=\bra{c_{\textrm{out}}}\widehat{\mathcal O}_{\textrm{out}}\ket{a_{\textrm{out}}}\;.
\ee

The elements of the density matrix for the outgoing radiation can be found by calculating different moments of the operators $\{b_i^+,\,b_i\}$. In particular, the number operator of the $j$th outgoing mode is
\be
\left\langle n_j\right\rangle\equiv\bra{0_-}b_j^+ b_j\ket{0_-}=\sum_a n_{ja}\,{\rho^{\textrm{out}}}_{aa}\;.
\ee

The famous result of Hawking is that these elements are the same as for thermal radiation,
\be\label{hawking}
{\rho^{\textrm{out}}}_{ac}=\prod_j \delta_{n_{ja}n_{jc}}P(n_{ja})\;,\qquad P(n)=\frac{(1-e^{-\omega/T})(e^{-\omega/T}\Gamma)^n} {\left[1-(1-\Gamma)e^{-\omega/T}\right]^{n+1}}\;,
\ee
where $\Gamma$ is the grey-body factor. The density matrix \eqref{hawking} is diagonal in both $j_a$ labeling the mode
frequencies and $n_{ja}$, the occupation number of each mode. The index $a$ (also $c$) runs through the set of basis vectors of the Fock space that is built by the operators $b_i^+$. Hence, the dimension of the density matrix is formally infinite.

The result of Hawking indicates that the density matrix for the outgoing radiation is thermal and thus is similar to that of a maximally mixed state. This is in contradiction to unitary evolution and at the core of the information paradox since the collapsing matter was in a pure state and evolved into a mixed state. Nonetheless, we will eventually show how the presence of off-diagonal elements in the density matrix $\widehat\rho^{\textrm{\; out}}$ can lead to the purification of the outgoing radiation in the course of the BH evaporation.


\section{The multi-particle density matrix in terms of the single-particle density matrix}\label{sec:denmat}

The goal of this section is to expose the connection of the full density matrix of the outgoing radiation $\widehat\rho^{\textrm{\; out}}$, as defined in \eqref{dm}, to the widely used single-particle density matrix
\be\label{rho1}
\rho^i_{\;j}=\bra{0_-} b^{+}_j b^i \ket{0_-}\;.
\ee

Let us begin with the well-known Bogolyubov transformation between the creation and annihilation operators at future infinity $\{b_i^+,\,b^i\}$ and at past infinity $\{a_i^+,\,a^i\}$. This is expressible as\footnote{Here, upper and lower indices have been introduced so as to make the matrix multiplication explicit. The
summation symbol will typically be omitted in what follows.}
\be\label{bogo}
b^i = \sum_j \l \bar\alpha^i_{\;j} a^j - \bar\beta^{ij} a^+_j \r \;, \qquad b^+_i = \sum_j \l \alpha_i^{\;j} a^+_j - \beta_{ij} a^j \r \;.
\ee
In terms of the Bogolyubov coefficients, we can express $\rho^i_{\;j}$ as follows:
\be\label{rho}
\rho^i_{\;j} =\sum_k \bar\beta^{ik}\beta_{jk} \;.
\ee

As shown in Appendix~\ref{sec:app_formula},  there is a closed expression for the density matrix $\widehat \rho^{\textrm{out}}$ in terms of the $b$'s,
\be\label{denmat}
\widehat\rho^{\;\textrm{out}} = \frac1Z e^{-b^+_i \Omega^i_{\;j} b^j} \;,
\ee
where $Z$ is the normalization factor and $\Omega$ is a $c$-number Hermitian matrix. The latter is related to the single-particle density matrix $\rho^i_{\;j}$ as
\be\label{omegarho}
\rho = \frac1{e^\Omega - 1} \;, \quad \text{or} \quad e^\Omega = 1 + \frac1\rho \;.
\ee
From now on, the superscript ``out'' on $\widehat\rho$ will be omitted for brevity,
\be
\widehat\rho \equiv\widehat\rho^{\;\textrm{out}}\;.
\ee

The full density matrix $\widehat \rho$ is thus completely defined by $\rho^i_{\;j} = \bar\beta^{ik}\beta_{jk}$, with the normalization factor $Z$ expressed as
\be\label{Z}
Z = \det \left[ \frac1{1-e^{-\Omega}}\right] = \det \left[ 1 + \rho \right] \;.
\ee

One can define the entropy and related quantities directly in terms of the matrix $\rho$. In particular, an explicit expression for the R\'enyi entropy, $H_2 \equiv - \ln \frac{\tr {\left[ \widehat\rho^{\; 2}\right]}}{\left(\tr\widehat\rho\right)^2} $, is readily obtained
by evaluating
\begin{align}
\tr {\left[ \widehat\rho^{\; 2}\right]} &= \frac1{Z^2} \tr\left[ e^{-2 b^+ \Omega\, b}\right] = \det\left[ 1-e^{-\Omega}\right]^2 \det \left[ \frac1{1 - e^{-2\Omega}} \right] = \nonumber\\
&= \det \left[ \frac{1-e^{-\Omega}}{1 + e^{-\Omega}} \right] = 
 \det \left[ \frac1{1 + 2\rho} \right] \;,
\end{align}
from which it follows that
\be\label{renyi}
H_2 = \tr{\left[ \ln{\l 1 + 2 \rho \r}\right]} \;.
\ee

From this expression, it would appear that the R\'enyi entropy does not vanish unless $\rho = 0$, when no particles have been emitted. However, the entropy for a state with $N$ radiated particles can reach as low as $H_2 = \ln{(2N)} \ll S_{BH}(0)$, which is below the scale of validity of the approximations that were used in deriving Eq.~\eqref{renyi}.

The expression for the von Neumann entropy is more cumbersome but we include it for completeness:
\begin{equation}
S = - \tr {\left[ \widehat\rho \ln \widehat\rho \right]}
= \tr\ln \left[ 1+\rho \right] + \det \left[ \rho \ln\l 1 + \frac1\rho\r\right] \;.
\end{equation}
In this paper, we will therefore concentrate on the R\'enyi entropy $H_2$. It is much easier to calculate and  provides an accurate measure of the entanglement in the state of the BH radiation.\footnote{For Gaussian states, such as the vacuum state of a non-interacting theory, $H_2$ respects the strong-subadditivity condition in the same way the von Neumann entropy does~\cite{Adesso:2012ni}.}


\section{The state of the emitted particles in the Hawking model}\label{sec:state}

Having a formal expression for the density matrix at hand, we can now evaluate the R\'enyi entropy and the particle number in the state of the outgoing radiation in the Hawking framework. However, as made clear below, these quantities are formally proportional to an infinite sum over all the possible modes in the Fock space of the outgoing radiation. This divergence comes about from implicitly assuming an eternal BH that radiates for an infinite period of time and, hence, emits an infinite amount of particles.

The physical situation, however, must be different: The number of particles emitted by the BH during a finite time interval is finite and depends on the initial mass of the BH. The goal of this section is to present a way of counting the modes emitted by the BH which reach an observer at future infinity during some finite time interval.\footnote{For an earlier, detailed calculation of the physical emission rates that employs  a different technique, see Ref.~\cite{Page:1976df}.} By implementing this method, we find that the physical quantities indeed become finite. For simplicity, the analysis is initially specialized
to the case of the Hawking model.

In the current case, the matrix $\Omega^i_{\;j} = \frac{\omega_i}{T} \delta^i_{\ j} $ is diagonal, and Eqs.~\eqref{Z} and~\eqref{renyi} for the partition function and the R\'enyi entropy then read
\begin{align}\label{z_hawk}
 - \ln Z &= \sum_i \ln \l 1- e^{-\omega_i/T}\r \;,\\
H_2 &= \sum_i \ln \l \frac{1+ e^{-\omega_i/T}}{1- e^{-\omega_i/T}}\r \;,
\end{align}
while the total particle number is given by
\begin{equation}\label{n_hawk}
N = \sum_i \frac1{e^{\omega_i/T} - 1} \;.
\end{equation}

The sums in the above expressions run over the full set of the frequency modes $\omega_i$ of a quantum field in the BH background. Hence, the dimension of the set of the possible frequencies is infinite, leading to an infinite entropy and particle number.

\subsection{A reminder: A scalar field in a box}

To better understand the correct way of counting the modes so as to render physical quantities finite, let us first recall a simple case: A massless quantum field in a box.

A massless quantum scalar field $\phi(t,\vec{x})$ in a box of size $L$ with reflecting boundary conditions
({\em i.e.}, $\phi(t,0)=\phi(t,L)=0$) can be expanded in terms of the Fourier modes
\be
u_{\vec{k}}(t, \vec{x}) \equiv \frac1{(2 \pi L)^{3/2}} \frac1{(2 \omega)^{1/2}}\sin{\l \frac{2\pi}{L} k_x x\r} \sin{\l \frac{2\pi}{L} k_y y\r} \sin{\l \frac{2\pi}{L} k_z z\r} e^{-i \omega t} \;,
\ee
with frequencies $\omega = \frac{2\pi}L \, \sqrt{k_x^2 + k_y^2 + k_z^2}$, where $k_x$, $k_y$ and $k_z$ can be any integer.

For a large enough box ($L \gg 1/T$), the sum over the modes can be approximated by an integral over a continuous momentum $\vec{p}$,
\begin{equation}\label{extensive}
\sum_{k_x, k_y, k_z} \simeq L^3 \int \frac{d^3 p}{(2 \pi)^3} \;,
\end{equation}
with the corresponding frequency given by $\omega(\vec{p}) = | \vec{p} |$. In this way, the quantities in Eqs.~\eqref{z_hawk}--\eqref{n_hawk} are proportional to the volume of the box $V= L^3$; that is, they are extensive. Moreover, since the only other dimensional parameter is the temperature $T$, any extensive, dimensionless quantity is given by $V \,T^3$ times some numerical constant:
\begin{align}
 - \ln Z & = V \, T^3 \cdot \frac1{2 \pi^2} \int_0^\infty dx \, x^2\ln \l 1- e^{- x}\r =-\frac{\pi^2}{90} V \, T^3 \;,\\
H_2 &= V \, T^3 \cdot \frac1{2 \pi^2} \int_0^\infty dx \, x^2 \ln \l \frac{1+ e^{-x}}{1- e^{-x}}\r =\frac{\pi^2}{48} V \, T^3\;,\\
N &= V \, T^3 \cdot \frac1{2 \pi^2} \int_0^\infty dx \frac{x^2}{e^x - 1}=\frac{\zeta(3)}{\pi^2} V \, T^3 \;,
\end{align}
where $x=\omega/T$.

In general, due to the diagonal form of the thermal density matrix, any quantity that is given by a trace over the Hilbert space $\mathcal H_{\textrm{out}}$ is proportional to a product over the available frequencies (as in Eq.~\eqref{norm}). A logarithm of such a quantity is then proportional to a sum over the frequencies and is extensive by virtue of Eq.~\eqref{extensive}. Imposing some physical boundary conditions on the scalar field results in a restriction on the allowed frequency modes---in this case, the condition that the wave numbers have to be integer multiples of $2\pi/L$. The formally infinite quantities \eqref{z_hawk} - \eqref{n_hawk} then become finite, as should be true for any physical system.

\subsection{Emission of localized wave packets}

 We have just seen that, for a field in a box, the frequencies are quantized in units of the inverse size of the box and the total number of modes is proportional to the volume of the box. In order to study how a BH radiates into an infinite space, it is more convenient to use localized wave packets with finite normalization instead of the Fourier modes (as in~\cite{Hawking:1974sw}).

Let us begin here with a scalar field that is propagating in the background of a BH. It can be expanded in terms of the complete set of solutions of the wave equation having only positive frequencies. At past infinity, the expansion contains only ingoing modes and takes the form
\be
\phi=\sum_i\left(f_ia_i+\bar f_i a_i^+\right)\;,
\ee
where $\{a_i^+,\,a_i\}$ are the creation and annihilation operators defined at past infinity and the sum runs over a discrete set of modes $\{f_i\}$ with finite normalization. At future infinity, the expansion of the scalar field in terms of positive-frequency solutions contains both ingoing and outgoing modes,
\be
\phi = \phi_{\textrm{in}}+\phi_{\textrm{out}}\;,\qquad \phi_{\textrm{out}}=\sum_i\left(p_ib_i+\bar p_ib_i^+\right)\;,
\ee
where $\{b_i^+,\,b_i\}$ are the creation and annihilation operators of the outgoing modes $\{p_i\}$ at future infinity.

 In the continuum normalization with frequencies $\omega$ (rather than the discrete wave packets $i$), the ingoing and outgoing Fourier modes for the spherically symmetric solutions are expressible as
\begin{align}
f_{\omega lm}(v, r, \theta, \phi) &= F_{\omega lm}(r) Y_l^m (\theta, \phi) e^{i \omega v} \;, \\
p_{\omega lm}(u, r, \theta, \phi) &= P_{\omega lm}(r) Y_l^m (\theta, \phi) e^{i \omega u} \;,
\end{align}
where $l$ and $m$ are the angular-momentum numbers, and $v$ and $u$ are the advanced and retarded tortoise coordinates (respectively).

A complete set of localized wave packets can then be defined as~\cite{Hawking:1974sw}
\begin{align}
f_{jnlm} (v, r, \theta, \phi) &= \eps^{-1/2} \int_{j \,\eps}^{(j+1) \,\eps} e^{-2\pi i \, n \, \omega/\eps} \,f_{\omega lm}(v, r, \theta, \phi) \, d\omega \;, \\
p_{jnlm} (u, r, \theta, \phi) &= \eps^{-1/2} \int_{j \,\eps}^{(j+1) \,\eps} e^{-2\pi i \, n \, \omega/\eps} \,p_{\omega lm}(u, r, \theta, \phi) \, d\omega \;.\label{packet}
\end{align}
In place of the continuous label $\omega$, the wave packets have  been labeled by two integer indices $j \ge 0$ and $n$. The wave packet with index $j$ contains waves with frequencies that are localized in the range $j\, \eps \le \omega \le (j+1) \, \eps$. Index $n$ labels the ray along which the packet is propagating: The wave packet $f_{jn}$ is peaked around the advanced time $v = 2\pi n \eps^{-1}$ and the wave packet $p_{jn}$, around the retarded time $u = 2\pi n \eps^{-1}$,
both having a width of $2\pi\eps^{-1}$.

Our particular interest is the wave packets which were emitted by the collapsing body and then detected at some fixed distance away from the BH during a given period of time $t_0 \le t \le t_0 + \Delta t$. Such wave packets will also be localized in the corresponding range of retarded time $u_0 \le u \le u_0 + \Delta t$. For a long enough time interval, one can always choose the width $\eps^{-1}$ so as to ensure the detection of many wave packets
($\eps^{-1} \ll \Delta t$), as well as maintain a fine-enough frequency resolution ($\eps \ll T$)
for each wave packet to be treated as a monochromatic mode of fixed frequency. In such a case, the summation over the wave-packet position $n$ in the interval $\Delta n = \Delta t/(2 \pi\eps^{-1})$ can be approximated as
\be
\sum_{n=n_0}^{n_0 + \Delta n} \approx \frac{\eps}{2\pi} \Delta t\;,
\ee
whereas the summation over the discrete set of frequencies $j=\omega/\eps$ can be approximated by the integral
\be
\sum_{j=0}^\infty \approx \int_0^\infty \frac{d\omega}{\eps} \;.
\ee

As one can now see, in this approximation, the total number of modes which can be detected during the time interval $\Delta t$, does not depend on the choice of the parameter $\eps$ and is proportional to $\Delta t$ times the integral over mode frequencies,
\be
\sum_i = \sum_{j=0}^\infty\sum_{n=n_0}^{n_0 + \Delta n}=\Delta t \, \int_0^{\infty} \frac{d \omega}{2 \pi} \;.
\ee

The total number of particles detected during the time interval $\Delta t$ in this case is given by
\be\label{number}
N = \sum_{j = 0}^\infty \sum_{n = n_0}^{n_0 + \Delta n} \sum_{l= 0}^\infty \sum_{m = -l}^l \frac{\Gamma_{j lm}}{e^{j \eps/T} -1}
= \Delta t \, \sum_{l= 0}^\infty \sum_{m = -l}^l \int_0^\infty \frac{d \omega}{2 \pi} \frac{\Gamma_{\omega lm}}{e^{\omega/T} -1} \;,
\ee
where $\Gamma_{\omega lm}$, the so-called grey-body factor, is determined by the properties of the modes near the horizon.

The emission of the modes with high multipoles is highly suppressed,
$\Gamma_{\omega lm} \ll \Gamma_{\omega 00} \equiv \Gamma(\omega)$ for $l ,m > 0$, and the sum~\eqref{number} can be well approximated with the $l = m = 0$ term only (see, {\em e.g.},~\cite{Page:1976df}). In what follows, we will consider only the $l=m=0$ modes and omit the angular-momentum labels.
Moreover, in the Schwarzschild case, there is only one available dimensional parameter, which can be chosen as the BH temperature $T$. Therefore, the grey-body factor $\Gamma(\omega)$ depends on the frequency only through the ratio $x = \omega/T$, and the number of emitted particles simplifies as follows:
\be\label{n_emit}
N = \Delta t \, T \int_0^\infty \frac{d x}{2 \pi} \frac{\Gamma(x)}{{e^x} -1} \sim \Delta t \,A \, T^3 \int_0^\infty \frac{d x}{2 \pi} \frac{\Gamma(x)}{{e^x} -1}\;,
\ee
where $A \sim T^{-2}$ is the area of the BH horizon.

 Thus, the rate of BH emission coincides with the thermal emission rate of a body with area $A$. In reality, due to the non-trivial frequency dependence of the grey-body factors, the spectrum for BH radiation is quite different from the thermal spectrum of a body in an empty box. Some details about the grey-body factors are provided
in Appendix \ref{sec:app_greybody}.

And so, in the absence of other dimensional parameters, all the extensive quantities describing the emitted radiation are proportional to the product $\Delta t \, T$ or, equivalently, to the total number of emitted
particles~\eqref{n_emit}. For example, the R\'enyi entropy $H_2$ is given by
\be\label{renyiH}
H_2 = \Delta t \int_0^\infty \frac{d \omega}{2 \pi} \ln{\l 1 + \frac{2\Gamma(\omega/T)}{e^{\omega/T} -1} \r} = \Delta t \, T \int_0^\infty \frac{d x}{2 \pi} \ln{\l 1 + \frac{2\Gamma(x)}{e^{x} -1} \r} \propto N\;.
\ee

Hence, by properly accounting for the finite duration of detection, one finds that the number of particles
emitted by the BH, as well as the entropy of the radiation, becomes finite as opposed to the idealized case of infinite emission time.


\subsection{Single-particle density matrix in the wave-packet basis}\label{sec:dm_wavepack}
In order to highlight the connection between the localized wave-packet description and the semiclassical model as  presented in~\cite{Brustein:2013qma,Brustein:2013ena},  we shall construct the single-particle density matrix $\rho^i_{\; j}$ in the finite wave-packet basis.

Using the definition of the wave packets \eqref{packet}, one can express the matrix elements of $\rho$  in terms of  the  same Fourier-mode basis,
\be\label{rhojn}
\rho^{j n}_{\quad j' n'} = \eps^{-1} \int_{j \eps}^{(j+1) \eps} d \omega \, e^{2\pi i \, n \, \omega/\eps} \int_{j' \eps}^{(j'+1) \eps} d \omega' \, e^{- 2\pi i \, n' \, \omega'/\eps} \rho(\omega, \omega') \;,
\ee
where
\be\label{singlerho}
\rho(\omega,\omega')=\int_0^\infty d\tilde\omega\,\bar\beta_{\omega\tilde\omega}\beta_{\omega'\tilde\omega}\;.
\ee

Each wave packet is labeled by the characteristic frequency $\omega_j = j \, \eps$ and the detection time $t_n=2\pi\eps^{-1}n$. The choice of $\eps$ parametrizes a trade-off between the frequency and the position resolutions. The total number of modes $W$,  within a given range of frequencies  $\Delta \omega$ and an
interval of time $\Delta t$,  does not depend on $\eps$ and is just given by $W=\Delta j\cdot\Delta n=\Delta t \, \Delta \omega / (2\pi)$. In the case of a thermal distribution, the frequency range $\Delta\omega$ is effectively set by the temperature. Hence, the total number of modes with substantial occupation numbers is of the order  $W=T \, \Delta t$. The matrix $\rho^{j n}_{\quad j' n'}$ in Eq.~\eqref{rhojn} is, therefore, effectively a $W \times W$ matrix.

An important feature of  BH radiation  is that the emission rate $\Gamma$ is the same  as the temperature $T$. For a general radiating body, this is not true; even if $\Gamma\propto T$, the proportionality constant need not be close to
unity. This has important consequences, which we now elaborate on.

 From the above discussion on  wave packets,  we have learned that the total number of emitted  modes  is given by $W = \Delta \omega  \Delta t$, where $\Delta \omega$ is the range of emitted frequencies and $\Delta t$
is the detection time. For the case of a nearly thermal
emitter, the frequency range is given by the  temperature  $\Delta \omega \sim T$, so that $W= T\Delta t$. On the other hand, the total number of emitted particles during this same time  is given by $N = \Gamma  \Delta t$, where $\Gamma$ is the emission rate. For a BH, with the emission rate $\Gamma\sim T$, the number of emitted particles $N$ is then of the same order as the total number of  occupied frequency modes $W$; {\em i.e.},  $N\sim W$. This fact has two important consequences for BH radiation:

First, the average occupation number for the modes of radiation is approximately of order unity, as  $N/W \sim \Gamma /T \sim 1$;  meaning that,
on average, each mode of radiation is occupied by only a few particles.
Second, since  $W\sim N$, the single-particle density matrix $\rho$ is effectively of  size $N\times N$.
Then, as  the average occupation numbers correspond to the diagonal elements of the matrix $\rho$,  these elements  go as
 $\rho^{j n}_{\quad j n} \sim 1$.  Hence, for BH radiation, the two-point function $\rho$ in the wave packet basis has the  form of an  $N \times N$ \emph{identity} matrix.

The fact that, on average, each  mode is occupied by only a few particles also implies that the radiation field in a BH background cannot be treated as semiclassical.
This is not the normal state of affairs.
 For any other thermal emitter, the radiation rate is proportional to its area,
$\Gamma \sim A \cdot T^3$, and the area for classical emitters  is much larger than the inverse temperature squared ({\em i.e.}, the square of the
typical
wavelength), so that  $\Gamma/T\sim A T^2\gg 1$.  Consequently, $N/W\sim \Gamma/T\gg 1$ and the
radiation field is highly classical.

In order to make these ideas  more precise, let us calculate $\tr\rho$ and
$\tr\rho^2$ for the Hawking model~\cite{Hawking:1976ra,Hawking:1974sw}. For a large BH,
\be\label{rhojnH}
\rho(\omega, \omega') = \delta(\omega - \omega') \, \rho(\omega) \equiv \delta(\omega - \omega') \, \frac{\Gamma(\omega)}{e^{\omega/T} - 1} \; .
\ee
With this expression, in the wave-packet basis,
\be
\rho^{j n}_{\quad j' n'} = \eps^{-1} \delta^j_{j'} \int_{j \eps}^{(j+1) \eps} d \omega \, e^{2\pi i \, (n - n') \, \omega/\eps} \, \rho(\omega) \;.
\ee

As one can observe, $\rho^{j n}_{\quad j' n'}$ depends only on the difference in  detection times $t_n - t_{n'}$. Because of the finite time and frequency resolution, it  has non-zero off-diagonal elements with $n\neq n'$, which are concentrated in a narrow strip $|n - n'| \lesssim \eps/T$  corresponding to $| t_n - t_{n'} | \lesssim T^{-1}$. The number of  non-zero elements of $\rho$ is, therefore, proportional to its dimension $W$ and, consequently,  of the order of the number of emitted particles $N$.

The traces of $\rho$ and $\rho^2$ can be explicitly calculated:
\be \label{tracerho}
\tr \rho = \sum_{j = 0}^\infty \sum_{n = n_0 + 1}^{n_0 + \Delta n} \rho^{j n}_{\quad j n} =
\eps^{-1} \sum_{n = n_0 + 1}^{n_0 + \Delta n} \int_0^\infty d \omega \, \rho(\omega) = \Delta t  \int_0^\infty \frac{d \omega}{2 \pi} \, \rho(\omega) \sim \Delta t \, T  = W\sim N\;,
\ee
\begin{align}
\tr \rho^2 &= \sum_{\substack{j = 0\\j'=0}}^\infty \sum_{\substack{n = n_0 + 1\\n' = n_0 + 1}}^{n_0 + \Delta n} \rho^{j n}_{\quad j' n'} \rho^{j' n'}_{\quad j n} \notag\\
&= \eps^{-2} \sum_{j = 0}^\infty \sum_{\substack{n = n_0 + 1\\n'=n_0 + 1}}^{n_0 + \Delta n}  \int_{j \eps}^{(j+1) \eps} d \omega  \int_{j \eps}^{(j+1) \eps} d \omega' \, e^{2\pi i \, (n - n') \, (\omega - \omega')/\eps} \rho(\omega)  \, \rho(\omega') \notag\\
&= \eps^{-2} \sum_{j = 0}^\infty \int_{j \eps}^{(j+1) \eps} d \omega  \int_{j \eps}^{(j+1) \eps} d \omega' \, \l \frac{\sin{[{\Delta n \, \pi \, (\omega - \omega')/\eps]}}}{\sin{[{\pi \, (\omega - \omega')/\eps]}}} \r^2 \rho(\omega)  \, \rho(\omega') \notag\\
&\approx \Delta t  \int_0^\infty \frac{d \omega}{2 \pi} \, \rho(\omega)^2 \sim \Delta t \, T \sim N \;.
\end{align}
In order to obtain the last line, we have used that $(\sin(\Delta n \pi x) / \sin(\pi x))^2 \approx \Delta n \,\delta(x)$ for large $\Delta n$.

The number of the non-zero eigenvalues, which  is just the number of diagonal elements in the Hawking model, can be then estimated by the participation ratio ({\em i.e.}, the inverse of the purity; see, {\em e.g.},~\cite{PR}),
\be\label{pr}
PR\equiv\frac{(\tr{[\rho]})^2}{\tr{[\rho^2]}} \sim N\;.
\ee
The result grows linearly with the number of emitted particles and also provides an estimate of the R\'enyi entropy~\eqref{renyiH}.

For a more general case including off-diagonal elements, one can still expect $\tr \rho\sim N$, as this result only
depends on the dimensionality of the matrix. On the other hand, $\tr \rho^2$ can be much different, but its value
can be readily estimated if the number of non-zero eigenvalues is known or {\em vice versa}.

\section{Off-diagonal elements of the density matrix}\label{sec:off_diagonal}

With the previous framework in hand, we will now consider the structure of the off-diagonal elements in both the single- and multi-particle density matrices. This will be carried out for the three distinct models of an evaporating BH:
(1) Hawking's model~\cite{Hawking:1974sw}, (2) the Page model~\cite{Page} and (3) the semiclassical model~\cite{Brustein:2013qma,Brustein:2013ena}.\footnote{For the Page model, the single-particle density matrix cannot be calculated without additional information because this framework is based upon choosing an arbitrary random basis in the combined radiation and BH Hilbert space. How this basis
is related to the Fock space of the radiation is left unspecified.}

In this section, all dimensional quantities are expressed in units of
the BH temperature, $N=N(t)$ is the total number of emitted
particles at time $t$ and $S_{BH}\equiv S_{BH}(N)=S_{BH}(0)-N(t)$ is the BH entropy at ``time'' $N$. This relation is modified, semiclassically, by fluctuations in the emission time of the Hawking modes. However, the fluctuations were shown to be small~\cite{Brustein:2013qma}, being proportional to inverse powers of $S_{BH}$.

One of the main issues to be clarified is to what extent the off-diagonal elements in terms of  mode-occupation number of the full density matrix are suppressed. If they are indeed highly suppressed, then the relationship between the density matrix $\widehat\rho$ and the single-particle density matrix $\rho$ is reliable.

\subsection{The Hawking model}\label{sec:off_Hawking}

In Hawking's model, the outgoing radiation is described to a very good approximation by the diagonal density matrix given in Eq.~\eqref{hawking}. What we would like to address here is how accurate an approximation this is by quantifying the off-diagonal contribution.\footnote{By off-diagonal contribution, we mean that beyond the broadening effects of the wave-packet treatment in the last section.} We shall, however, not discuss any back-reaction nor evolution effects to the Hawking model in this subsection. These effects were originally disregarded by Hawking himself and introducing them would mean deviating from the original model in a significant way. We do account for the evolution effects in the semiclassical model that we discuss in the subsection~\ref{sec:scmodel}.

The Hawking density matrix $ \widehat{\rho}_H$ is only approximately diagonal in two obvious ways: Elements between states differing in total occupation number are only approximately vanishing, as are those between states with some modes differing in frequency or in the time of emission. We would like to estimate the magnitude of the off-diagonal elements of the density matrix. As far as we know, this issue has not been addressed directly in the literature, although Hawking did expect such corrections to be exponentially small~\cite{Hawking:1974sw}.

What we  find below is an overall suppression for off-diagonal elements in mode-occupation number that is stronger than $e^{-S_{BH}}$. This is significant  because it implies that the results of Hawking are not modified in a meaningful way by off-diagonal corrections.\footnote{See, however, the discussion in the Introduction about corrections arising from geometries without a horizon.} We also find that the off-diagonal elements of the single-particle density matrix $\rho^i_{\ j}$ are smaller than $1/S_{BH}$, leading to corrections to physical quantities, such as the entropy, that are suppressed by at least a factor of $1/S_{BH}^2$. These are negligible at all times, as originally anticipated by Hawking.  Corrections to the diagonal elements of $\rho^i_{\ j}$ are similarly suppressed by  at least $1/S_{BH}$ and  readily absorbed into the normalization.

The fact that the Hawking density matrix is diagonal can be understood by looking at Hawking's derivation of the single-particle density matrix $\rho_H$, as  defined in Eq.~\eqref{singlerho}.  The relevant factor goes as  $\rho_H(\omega,\omega') \sim I_{-}(\omega,\omega')$ such that  ~\cite{Hawking:1974sw}
\be
I_{-}(\omega,\omega') =
\int\limits_{0}^{\infty} d\tilde{\omega} \; (\tilde{\omega})^{-1+ i(\omega-\omega')}\;  = \int\limits_{-\infty}^{\infty} dy \; e^{+iy(\omega-\omega')}
= 2\pi\,\delta(\omega-\omega')\;,
\label{freqint}
\ee
where $y=\ln{\tilde{\omega}}$ has been  used. This equation makes it clear
 that the matrix elements of $\rho_H$ are diagonal in frequency space.

The off-diagonal elements of $\widehat\rho_H$ for different occupation numbers
are also vanishing, as follows from an estimation of  the products
\be\label{abcX}
\alpha\beta\equiv\int_0^\infty d\tilde\omega\,\alpha_{\omega\tilde\omega}\beta_{\omega'\tilde\omega}\;,\qquad \bar\alpha\bar\beta\equiv\int_0^\infty d\tilde\omega\,\bar\alpha_{\omega\tilde\omega}\bar\beta_{\omega'\tilde\omega}\;.
\ee
The products $\alpha\beta$ and $\bar\alpha\bar\beta$ arise in the expectation values of operators like
\begin{align}\label{k1}
&\bra{0_-}\left(b^+\right)^nb^{n+2k}\ket{0_-}\sim \left(\beta\bar\beta\right)^n\cdot\left(\alpha\beta\right)^{k}\;,\\\label{k2}
&\bra{0_-}\left(b^+\right)^{n+2k}b^{n}\ket{0_-}\sim \left(\beta\bar\beta\right)^n\cdot\left(\bar\alpha\bar\beta\right)^{k}\;
\end{align}
and, thus, define the elements of the density matrix between states that differ in total occupation number by $\Delta {\cal N}=2k$. The products $\alpha\beta$ and $\bar\alpha\bar\beta$ also appear in the computation of the generating function \eqref{gen_fun_1}. Consequently, the dependence of the density matrix on $\rho$ is only truly valid in the case when the products $\alpha\beta$ and $\bar\alpha\bar\beta$ are negligible.

In Hawking's Fourier-space analysis, such terms contain an integral that is  similar to the one in Eq.~\eqref{freqint} but now the two frequencies appear with the same sign,
\be
I_{+}(\omega,\omega') =
\int\limits_{0}^{\infty} d\tilde{\omega} \; (\tilde{\omega})^{-1+ i(\omega+\omega')}\;  = \int\limits_{-\infty}^{\infty} dy \; e^{iy(\omega+\omega')}
= 2\pi\,\delta(\omega+\omega')\;.
\label{freqint1}
\ee
Since both frequencies are positive, the argument of the delta function can only vanish  if both of the frequencies vanish. This leads to one or more  factors of $\delta(\omega+\omega')$ when contracting pairs of operators in the corresponding density-matrix element, as there would necessarily be unequal numbers of creation and annihilation operators within the expectation value.

To help reveal where corrections to off-diagonal elements can appear, let us consider the integration limits in Eqs.~(\ref{freqint}) and~(\ref{freqint1}). These limits are an idealization, as a frequency can never be truly infinite nor exactly zero in physically realistic situations. For instance, the upper limit on $\tilde{\omega}$  should be an exponentially large number, representing ultra-high frequencies on  past null infinity that need to be red-shifted in order to produce the thermal radiation at future null infinity.  An estimate of this upper limit in units of the BH mass $M$ is $\sim e^{\tau/M} \sim e^{(M/M_{Pl})^2}\sim e^{S_{BH}(0)}$, where $\tau\sim M^3$ denotes the lifetime of the BH (see, {\em e.g.},~\cite{Ford:1997hb}).  As for the lower limit on $\tilde{\omega}$, a natural choice is to fix it as the inverse of the BH lifetime $(\tau/M)^{-1}\sim S^{-1}_{BH}(0)$. However, in Hawking's model, the BH
is regarded as eternal, and so this limit should rather be the exponential of a large negative number.

In the following, we replace the idealized limits with suitable ultraviolet and infrared cutoffs, $e^{y_{max}}$ and $e^{-y_{min}}$, which are assumed to be of order $y_{max}\sim y_{min}\sim S_{BH}(0)$. We will further set $y_{\ast}\equiv y_{max}=y_{min}$, as this symmetry will simplify the calculations without affecting the conclusions. This estimate determines the strength of the off-diagonal corrections of the density matrix in mode-occupation number. If, for some reason, the limits are such that $y_{max}\sim y_{min}\ll S_{BH}(0)$, then the original Hawking calculation is significantly modified and the original conclusion about the nature of the density matrix needs to be revised.

The new choice of limits leads to a correction to the previous delta function,
\bea
\int\limits_{-y_{\ast}}^{y_{\ast}} dy \; e^{iy(\omega\pm\omega')} &=2y_{\ast}\frac{\sin{(y_{\ast} (\omega\pm\omega'))}}{y_{\ast}(\omega\pm\omega')}\;.
\eea
The tails of this ``regulated'' delta function
provide the off-diagonal correction of interest. Notice that this comes about for both $I_\pm(\omega,\omega')$, and, hence, the off-diagonal elements of the density matrix can now be non-vanishing, even
when connecting  states that  differ in total occupation number.

To better understand the implications of these off-diagonal corrections,  the single-particle  density matrix (and its related products) can be convolved with the same wave packets as in Eq.~(\ref{rhojn}). This analysis is carried out in detail in Appendix~\ref{sec:offf_diagonal}, from which  there are two main conclusions: First, we find that the off-diagonal elements of the \emph{single-particle} density matrix $\rho^{j n}_{\quad j' n'}\equiv (\bar\beta\beta)^{j n}_{\quad j' n'}$ are suppressed relative to the diagonal elements by a factor of $\mu^{-2}_* J^{-1}$, where $\mu_*\equiv y_*\eps/(4\pi^2 T)$ and $J=j'-j$. Second, it is found that all the elements of the products $\alpha\beta,\,\bar\alpha\bar\beta$ are subleading and  suppressed as $\mu^{-2}_* J^{-1}_{\alpha\beta}$. Here, $J_{\alpha\beta}=j'+j+1$ and can be treated as some typical average value of the frequency label $j$ referring to the frequency range $\omega_j\in\left[j\eps,(j+1)\eps\right]$. These products set the order of magnitude of the off-diagonal elements of the \emph{full} density matrix. In particular, this result tells us that the leading-order contribution to a typical off-diagonal element of the density matrix\footnote{Typical in the sense that only an exponentially small fraction of the off-diagonal elements deviate from this behavior.} is relatively suppressed by a factor of $\mu_{\ast}^{-|\Delta {\cal N}|}j_{\ast}^{-\frac{1}{2}|\Delta {\cal N}|}$. Here $\Delta {\cal N}$ is the difference in total occupation number of the corresponding states and $j_{\ast}$ can be viewed as a typical value for the integer $j$.

Let us elaborate on this last claim by, first, understanding the structure of the density matrix $\hat\rho$. A natural way of organizing the  elements of the density matrix which connect  different Fock states is to classify them by
 the total number of particles in each state. A density-matrix element ${\widehat{\rho}^{\cal N}}_{\;\;\cal N'}$ then denotes a block of elements connecting states with total occupation
numbers
 ${\cal N}$ and ${\cal N}'$. Thus, each element ${\widehat{\rho}^{\cal N}}_{\;\;\cal N'}$ is  a matrix of dimension $d_{\mathcal N'}\times d_{\mathcal N}$, where $d_{\mathcal N}$ is given by the size of the Fock subspace spanned by the states with total occupation number equal to $\mathcal N$. If we denote the number of available frequency modes by $W$, then $d_{\mathcal N}$ coincides with the number of distinct elements in a symmetric tensor of rank $\mathcal N$ in a $W$-dimensional vector space:
 \be\label{dn}
  d_{\mathcal N}= \begin{pmatrix}
\mathcal N+W-1\\\cal N
\end{pmatrix}\;.
\ee

Suppose that our interest is the
strength of the density-matrix element $\widehat{\rho}^{{\cal N}}_{\;\;{\cal N}'}$ such that $\Delta {\cal N}={\cal N}-{\cal N}'$ is an even, non-zero integer.
(There are
no contributions when this difference is odd; {\em cf}, Eqs.~\eqref{k1} and~\eqref{k2}.)
 The strength of such an element is determined by a higher-order moment than the two-point function;
rather, by an $({\cal N}+{\cal N}')$-point function with ${\cal N}$ annihilation operators and ${\cal N}'$ creation
operators. But, because the
theory of interest is non-interacting, the relative strength of such an
element can be assessed by looking
at products of two-point functions $\bar\beta\beta,\,\alpha\beta,\,\bar\alpha\bar\beta$. Here, we will be able to correctly match a creation operator
with an annihilation operator for all but $\frac{|\Delta {\cal N}|}{2}$ of the total number of operator pairs
$\frac{{\cal N}+{\cal N'}}{2}$. Each mismatched pair---two creation operators or two annihilation operators---will then contribute one
suppression factor of $\mu^{-2}_* j_{\ast}^{-1}$.

Formally, the density matrix can be represented in terms of the products $\beta\bar\beta$, $\alpha\beta$, and $\bar\alpha\bar\beta$ as
\begin{align}\label{hierarchy}
\hat \rho\sim\quad
\begin{matrix}
\mathcal{ N'} = 1\rightarrow\\
&\\
\mathcal N'=2\rightarrow\\
&\\
\mathcal N'=3 \rightarrow\\
&\\
\mathcal N'=4\rightarrow\\
&\\
\vdots
&
\end{matrix}
&\begin{pmatrix}
\beta\bar\beta\;\;\;&0&\bar\alpha\bar\beta\;\;\;\;&0&(\bar\alpha\bar\beta)^2&0&\hdots\\
&&&&&\\
0&\beta\bar\beta\;\;\;&0&\bar\alpha\bar\beta&0&(\bar\alpha\bar\beta)^2&\hdots\\
&&&&&\\
\alpha\beta\;\;\;&0&\beta\bar\beta\;\;\;\;&0&\bar\alpha\bar\beta&0&\hdots\\
&&&&&\\
0&\alpha\beta\;\;\;&0&\beta\bar\beta&0&\bar\alpha\bar\beta&\hdots\\
&&&&&\\
\vdots&\ddots&\ddots&\ddots&\ddots&\ddots&\ddots
\end{pmatrix}\;.\\
&\quad\;\;\uparrow\qquad\uparrow\qquad\uparrow\quad\;\;\;\uparrow\quad\;\;\uparrow\qquad\;\uparrow\nonumber\\
&\mathcal N=1,\quad\,2,\quad\;3,\quad\;\;4,\quad\;5,\quad\;\;\,6 \quad \cdots \nonumber
\end{align}
Hence, a typical density matrix element between states with a difference in the total occupation number $\Delta\mathcal N$ is suppressed by the product of $\frac{\Delta{\cal N}}{2}$ factors of $\mu^{-2}_* j_{\ast}^{-1}$.

As  already stated, the off-diagonal contribution to $\rho_H$---the \emph{single-particle} density matrix--- is suppressed by a factor $\mu_*^{-2}(j-j')^{-1}\sim (y_*\eps/T)^{-2}(j-j')^{-1}$. Since the final result, when traced over all the frequency modes, should not depend on the parameter $\eps$, we can
choose it to be of the order of the BH temperature, $\eps\sim T$. This amounts to saying that all the modes have the same frequency and, hence, $j\sim\mathcal O(1)$. It is then clear that the off-diagonal contribution to $\rho_H$ is suppressed by at least $y_{\ast}^{-2}\sim S^{-2}_{BH}(0)< N^{-2}$ with respect to the diagonal elements and is, therefore, inconsequential to the moments of $\rho_H$.

Nevertheless, the question of whether the off-diagonal elements can make a substantial correction to the moments of the \emph{full} Hawking density matrix $\widehat{\rho}_H$ is well founded, as  the number of  off-diagonal elements is certainly  very large. To address this concern, let us consider the regime in which $N\sim S_{BH}$ because this is when significant changes to the nature of the
BH radiation could be expected. To estimate the importance of the corrections for the R\'enyi entropy $\frac{\textrm{tr}\widehat\rho_H^2}{(\textrm{tr}\widehat\rho_H)^2}$ we shall replace the hierarchical structure of the density matrix~\eqref{hierarchy} by a uniform estimate of the typical value of the off-diagonal corrections. We do so by setting $|\Delta {\cal N}|\sim N\sim S_{BH}$ for all the off-diagonal elements, so that the relative suppression factor goes as $(\mu_*^2j_*)^{-\frac{N}{2}}$. The overall off-diagonal contribution to the R\'enyi entropy can now  be calculated as the product of the dimensionality of the matrix $d$ and the square of the relative suppression factor, $d\times (\mu_*^2j_*)^{-N}$.

The size $d$ of the full density matrix for $N$ emitted particles is given by
\be
d= \sum_{n=1}^Nd_n=\sum_{n=1}^N\begin{pmatrix}
n+W-1\\n
\end{pmatrix}=\begin{pmatrix}
N+W\\ N
\end{pmatrix},
\ee
where $d_n$ is the size of the $n$-particle subspace defined in Eq.~\eqref{dn}, and $W$ is the number of available frequencies. Since  $W\sim N$ for BH radiation (see Subsection~\ref{sec:dm_wavepack}),
 this gives $d\sim e^{2N}$.

Thus, the overall off-diagonal contribution, relative to that of the diagonal, is  $e^{2N}(y_*^2)^{-N}\sim e^{2N}N^{-2N}\ll e^{-N}\sim e^{-S_{BH}}$, where we have again used  that $\eps\sim T$ and $j_*\sim \mathcal O(1)$, leading to the estimate  $\mu_*^2j_*\sim y_*^{2}\sim S_{BH}^{2}\sim N^{2}$. It can be concluded that the off-diagonal contribution to the Hawking density matrix is exponentially suppressed. This conclusion agrees with the analysis of
Mathur \cite{Mathur:2009hf}, who showed that small corrections at the level of
 the two-point function
cannot change any of Hawking's basic outcomes.  The physical reason behind this result is that the thermal radiation at future null infinity  originates from  ultra-high frequency modes at past null infinity which are  highly red-shifted  due to the presence of the horizon. We can therefore deduce that the existence of a region of high redshift---the horizon---induces a characteristic exponential suppression of the off-diagonal elements in mode-occupation number of the full density matrix.

Finally, let us use Eqs.~(\ref{renyi}) and~(\ref{tracerho}), to determine the R\'enyi entropy for the Hawking model. As  off-diagonal corrections are suppressed at least by order $e^{- S_{BH}}$ at all times and the diagonal is approximately uniform, the single-particle density matrix always has $N$ non-zero eigenvalues, each of which is unity up to negligible corrections. Hence, the estimate
\be
(H_2)_{\rm Hawking} \simeq N\ln{3}
\label{H2H}
\ee
is valid at all times.

\subsection{The Page model}

In the Page model of BH evaporation~\cite{Page}, the BH and the emitted radiation are assumed to be in a pure state in some large Hilbert space $\mathcal H$. This space is partitioned into an ``in" part, $\mathcal H_{BH}$, representing the BH (including the ingoing radiation)   and  an ``out" part,  $\mathcal H_{\textrm{out}}$, representing the outgoing Hawking radiation, so that $\mathcal H=\mathcal H_{\textrm{BH}}\otimes\mathcal H_{\textrm{out}}$. The two Hilbert spaces are characterized only by their dimensionality.

Let us label the states in $\mathcal H_{\textrm{BH}}$  by $i =1,\dots,m$, where  $m=e^{S_{BH}}$, and the states in  $H_{\textrm{out}}$  by $A=1,\dots,n$, where $n=e^{S_{rad}}$. A state in $\mathcal H$ can then be written as $\ket{A,i}$.
The working assumption is that both Hilbert spaces are large, $n,m\gg1$. Therefore, the Bekenstein--Hawking entropy $S_{BH}$ of the BH  and the radiation entropy  $S_{rad} \sim N$ (which is calculated as if the radiation was  in a thermal state)  characterize the dimensionalities of the corresponding Hilbert spaces rather than the real entropies of the subsystems. Since the total state is pure, the actual entropy of the radiation and that of the
BH are entirely due to entanglement and, thus, equal to one another at all times.

The partition between  ``in" and ``out" is meant to mimic the horizon separating the interior of the BH from its outside. The variation in time of the dimensions of the Hilbert spaces $\mathcal H_{\textrm{BH}}$ and $\mathcal H_{\textrm{out}}$  is meant to model the evaporation of the BH, so that $\mathcal H_{\textrm{BH}}$ shrinks and $\mathcal H_{\textrm{out}}$ grows while the number of emitted particles increases. However, additional physical effects resulting from the existence of a horizon are not taken into account.

The density matrix of the Page model $\widehat\rho_P$ is given by
\be
\widehat\rho_P = \rho_{A,i;B,j} \ket{A,i}\bra{B,j} \;.
\ee
The reduced density matrix for the radiation is then obtained by tracing over the BH Hilbert space, just as in Eq.~\eqref{dm},
\be
\widehat\rho_{\textrm{out}} = \tr_{BH}\ \widehat\rho_{P} \; .
\ee

Because the combined state is pure, there must be a special basis in which $\widehat\rho_P$ has only a single eigenvalue. This  can be chosen, without loss of generality, to be given by $\rho_{1,1;1,1}$. It follows that the matrix elements $\rho_{A,i;B,j}$ can be expressed in terms of a basis transformation matrix $U$ acting
 on the original state vector $\ket{1,1}$. Denoting  the resulting vector  as $V_{A,i}=U_{A,i;1,1}$, we have
\be
\rho_{A,i;B,j}= U_{A,i;A',i'}\;\rho_{A',i';B',j'}{U^{\dagger}}_{B',i';B,i}=V^{\phantom{+}}_{A,i} V^{{\ast}}_{B,j} \;
\ee
and
\be
\bra{A} \rho_{\textrm{out}}\ket{B} = \sum_{i} V^{\phantom{*}}_{A,i} V^{{\ast}}_{B,i} \;.
\ee
The vectors $V_{A,i}$ are meant to be treated statistically and assumed to be random vectors of unit size with a uniform distribution on an $m n$-dimensional sphere.

The initial investigation of such a random system was conducted by Lubkin~\cite{Lubkin} and then improved by many subsequent investigations~\cite{mp0,mp1,mp2}. The final result is that the distribution of eigenvalues of $\widehat\rho_{\text{out}}$ was found to obey the Marchenko--Pastur (MP) law~\cite{MPdist}, which is the eigenvalue distribution for a certain ensemble of semi-positive-definite,
 random matrices.

The MP distribution is generically divided into two parts, one set of large eigenvalues and another set of vanishingly small eigenvalues. It depends on only  two parameters, the total number of the eigenvalues and the participation ratio
corresponding to  the number of large eigenvalues. The distribution of the eigenvalues (up to a normalization convention) is given by
\begin{align}
\label{MP}
P_{MP}(\lambda) \, d\lambda &= \left( \max(1 - c, 0)\, \delta(\lambda) + \frac{c}{2\pi\lambda} \sqrt{(\lambda_+ - \lambda)(\lambda - \lambda_-) } \right) d \lambda \notag
\\ &= \left( \max(1 - c, 0)\, \delta(\lambda) + \frac{c}{2\pi\lambda} \sqrt{4 / c - (\lambda - (1 + 1/c))^2} \right) d \lambda \;,
\end{align}
where
\be
\lambda_+=(1+1/\sqrt{c})^2, \quad \lambda_-=(1-1/\sqrt{c})^2,
\ee
and $c > 0$ is a parameter that determines the number of large eigenvalues.

 When $c \gg 1$, all the eigenvalues are about equal and there are no zero eigenvalues whereas, for $c < 1$, a fraction $1-c$ of the eigenvalues is vanishing.
For a matrix of dimension $n$ with its eigenvalues distributed according to the MP law, the participation ratio PR, as defined in Eq.~\eqref{pr},
is given by
\be
PR \;= \frac{nc}{1+c} \;.
\label{MPPR}
\ee
The higher moments of the MP distribution are also known in terms of $n$ and
 $c$ (see Appendix~\ref{sec:moments}).

For the Page model, in particular,
\be
c = \frac{m}{n} \;,
\ee
and so
\be
PR=\frac{m n}{m+n} \;.
\ee

The early times of BH evaporation, when $n \ll m$, correspond to $c\gg 1$ and
 $PR \sim n$. Essentially, all the eigenvalues are about equal and their value is determined by the normalization convention. This means that the R\'enyi entropy of the radiation is the same as that of the Hawking model, $(H_2)_{\rm Page}\sim N \ln 3$. But well after the Page time,  when $n \gg m$, one finds that $c \ll 1$  and $PR \sim m \ll n$. In this case,  about $n - m \sim n$ of the eigenvalues become zero and both reduced density matrices for the radiation and  BH  are becoming pure.

The R\'enyi entropy of the radiation for the Page model coincides with that of BH and is given by
\be
H_{2} = \ln PR \simeq \begin{cases}
\ln n =S_{rad}\sim N & n < m \; ,\\
\ln m = S_{BH}(N)=S_{BH}(0) -N & n > m \;.
\end{cases}
\label{PageRenyi}
\ee
This is depicted in Figure~\ref{fig:MP} as a function of the number of emitted particles. Since the entropy of the subsystems is entirely due to entanglement, it is symmetric if one exchanges the Hilbert spaces of the BH and radiation. As a consequence the evolution of the entropy is symmetric under reflections around the Page time, when the sizes of $\mathcal H_{\textrm{BH}}$ and $\mathcal H_{\textrm{out}}$ are equal; that is, when half of the particles have been emitted, $N_{Page} \approx S_{BH}(0)/2$.

\subsection{The semiclassical model}\label{sec:scmodel}

The semiclassical model~\cite{Brustein:2013qma,Brustein:2013ena} improves upon Hawking's framework by taking into account the BH's quantum fluctuations, as well as its time dependence due to the emission of the radiation. In this setup, the single-particle density matrix $\rho_{SC}$ is no longer diagonal and, as a result, the evaporation process becomes unitary even though the thermal-like emission spectrum is kept.

The basic prescription is to assign a quantum wavefunction to the collapsing shell of matter in Hawking's model  and then recalculate all relevant quantities as expectation values. The main outcome is that $\rho_{SC}$ picks up off-diagonal contributions that are uniform in terms of frequency but suppressed relative to the diagonal elements by $\cbh^{1/2}(N)$, where $\cbh(N)=S_{BH}^{-1}(N)\ll 1$ is a classicality parameter---a ``time-dependent $\hbar$'' that keeps track of how close the system is to the limit of a classical spacetime.

The elements of $\rho_{SC}$ do have a non-uniform suppression in terms of emission
time; modes emitted at different times tend to decohere. Nonetheless, if the radiation
is being regularly monitored at intervals of $\Delta N\sim \sqrt{S_{BH}}$ or less, then this suppression can be compensated. We will specifically be considering this case, which has been called the ``tracking case'' in~\cite{endgame}. Hence, the
off-diagonal elements of the single-particle density matrix $\rho_{SC}$ can be regarded as uniform in magnitude with respect to both frequency and emission time.

As explained in Subsection~\ref{sec:dm_wavepack}, $\rho$ can be viewed as an $N\times N$ matrix, with the indices running over the wave-packet modes with non-vanishing occupation number and with the diagonal elements given by the average occupation number for each mode. The elements of the semiclassical single-particle density matrix are found to be~\cite{Brustein:2013qma,Brustein:2013ena}
\begin{align}\label{rhosc}
&(\rho_{SC})_{ii} = 1 \; , \notag \\
&(\rho_{SC})_{i\neq j} = \sqrt{C_\text{BH}(N)} e^{i \theta_{ij}} \; ,
\end{align}
where the phases $\theta_{ij}$ can be treated as random for most purposes (but see below).

Let us next consider the full density matrix $\widehat{\rho}_{SC}$ for this model. In ~\cite{Brustein:2013qma} (see, in particular, Sect.~2.4), it was argued that the off-diagonal elements in mode-occupation number of the density matrix  will be  suppressed much in the same way as for the Hawking model, as discussed in Subsection~\ref{sec:off_Hawking}. The physical reason for this is the same as before---the presence of the horizon implies that the frequencies observed at future null infinity are highly red-shifted and, thus, determine the magnitude of the suppression factor of the off-diagonal elements. The technical explanation is as follows: In the semiclassical model, the terms that lead to off-diagonal elements in mode-occupation number are, again, terms of the form $\alpha \beta$ and $\bar\alpha \bar\beta$ as in Eq.~(\ref{abcX}). These contain a classical term,
which is equal to the Hawking-model contribution, along with a semiclassical correction.  Irrespective of the details, it is clear that the semiclassical corrections to  $\alpha \beta$ and $\bar\alpha \bar\beta$ vanish as some power of $\cbh$ in the limit $\cbh\to 0$. This is enough to guaranty the exponential suppression of the semiclassical contribution to the off-diagonal elements in mode-occupation number.

Hence, there will be a hierarchical structure in the suppression factors of the off-diagonal elements in mode-occupation number, as shown in \eqref{hierarchy}. Like before, the off-diagonal elements can be expressed in terms of $({\cal N}+{\cal N}')$-point functions, which will factorize into a product of $\frac{{\cal N}+{\cal N}'}{2}$ two-point functions. The strength of an $({\cal N}+{\cal N}')$-point function will then be suppressed by a factor similar to the suppression factor of the Hawking model.

In some sense, this semiclassical model can be viewed as the ``middle ground'' between the other two models; on one hand, remaining almost thermal like Hawking's with the associated  hierarchical structure of the density matrix but, on the other,  evolving over time like Page's.  At the early stages of BH evaporation, the semiclassical model is essentially  Hawking's plus small corrections. However, at later times (after the Page time),  the dominant contributions will come
from those elements that are off-diagonal in frequency. This can be attributed to the effective perturbative parameter being $N\cbh$ for this  framework~\cite{Brustein:2013qma}, as
this parameter grows monotonically throughout the evaporation process and finally becomes large
($N\cbh>1$) at times later than  the Page time. It is the large size of
this effective perturbation parameter that allows the semiclassical
model to evade the conclusions of Mathur \cite{Mathur:2009hf}; in particular,
it leads to a  R\'enyi entropy for the radiation  that is decreasing in time (see below).

\subsubsection{Semiclassical R\'enyi entropy}

The objective here is to estimate the R\'enyi entropy of the BH radiation $H_2(N)$ using the single-particle density matrix $\rho_{SC}$ as specified in Eq.~\eqref{rhosc}. However, we are  currently lacking a precise quantitative knowledge of $\rho_{SC}$. In particular, in order to use Eq.~\eqref{renyi} for $H_2(N)$, one needs to know the eigenvalue distribution $P(\lambda)$ of $\rho_{SC}$:
\be\label{renyi_dist}
H_2(N) = \langle \tr{[ \ln{(1 + 2 \rho)} ]} \rangle_{P(\lambda)} = N \int_0^\infty \ln{(1 + 2 \lambda)} P(\lambda) d \lambda \;.
\ee
This deficiency  can be attributed to our lack of specification as to the exact nature of the phases $e^{i\theta_{ij}}$. These phases are not entirely random but rather dictated by the effect of the quantum horizon on correlations in the emitted radiation~\cite{Brustein:2013ena}. In this subsection, we
will instead call upon  the known properties of $\rho_{SC}$ in order to draw  conclusions about the entropy of the radiation in the semiclassical model.

We expect that $\rho_{SC}$ has some number of large eigenvalues with the rest vanishing, whereby the large eigenvalues are distributed about the mean in a way that depends on their total number. This is most clearly seen from the participation ratio $PR$ of the single-particle density matrix  $\rho_{SC}$. The participation ratio provides, for a given matrix, a measure of the number of non-vanishing eigenvalues ({\em e.g.}, \cite{PR}). For $\rho_{SC}$, it is given by~\cite{endgame}
\be\label{prsc}
PR \;= \;\frac{(\tr{\rho_{SC}})^2}{\tr{\rho_{SC}^2}} = \frac {N} {1 + N \cbh} = \frac {N(S_{BH}(0)-N)} {S_{BH}(0)} \;.
\ee
Then, since $\tr{\rho_{SC}} = N$ (as  shown in Subsection \ref{sec:dm_wavepack}),
the average value of a non-vanishing eigenvalue can be estimated as
\be
\overline{\lambda} \;=\frac{\text{tr }\rho_{SC}}{PR}=\;\frac{ N} {PR} = 1 + N \cbh \;.
\label{average}
\ee

This finding reveals  that, up to the Page time ($N \cbh < 1$),  there are about $N$ eigenvalues of order one. The properties of the emitted radiation in this regime coincide with the predictions of the Hawking model: Each of the $N$ available modes has an occupation number of order one and the radiation entropy grows linearly with the number of emitted particles.

On the other hand,  after the Page time ($N \cbh > 1$), the participation ratio~\eqref{prsc} stops growing and eventually becomes much smaller than $N$, so that the average $\overline{\lambda}$ is much larger than one. This signals that the matrix $\rho_{SC}$ has a small fraction of large eigenvalues, with the majority of eigenvalues vanishing. Physically, it means that, out of the $N$ available radiation modes, there is only a small fraction with large occupation numbers. In this regime, the entropy of the radiation~\eqref{renyi_dist} is dominated by
the contribution from the occupied modes and is smaller than the entropy of the thermal radiation~\eqref{H2H}  which is predicted by the original Hawking model:
\be
H_2(N) \simeq PR \, \ln{\left(1 + 2 \overline\lambda \right)} \sim PR \, \ln{\left( \frac{N}{PR} \right)} \ll N\;, \;\;\;\;\;\;{\rm for}\;\;PR\ll N\;.
\ee
Therefore, the known properties of the single-particle density matrix $\rho_{SC}$ guarantee that the radiation entropy in the semiclassical model does not grow steadily with number of emitted particles, as predicted by the Hawking computation, but rather starts to decrease after the Page time.

\begin{figure}
\centering
\includegraphics[width = .7\linewidth]{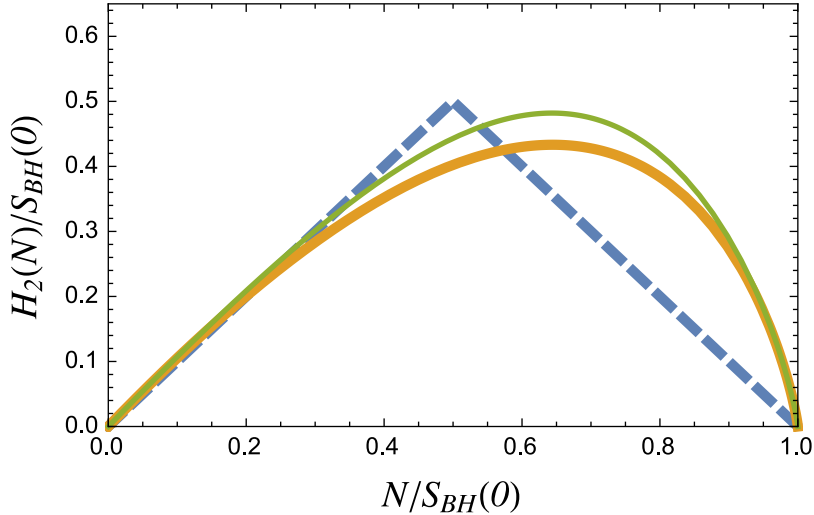}
\caption{Dependence of the R\'enyi entropy $H_2$ of the BH radiation on the  number of emitted particles $N$ for the semiclassical model in Eq.(\ref{rhosc}) for two eigenvalue distributions. First, assuming that the non-vanishing eigenvalues of the two-point function are distributed according to Eq.~\eqref{dist_equal} (solid, thick orange) and then to the Marchenko--Pastur law in Eq.~(\ref{MP}) (solid, thin green). For comparison, the dashed blue line depicts the prediction of the Page model from Eq.~(\ref{PageRenyi}).\label{fig:MP}}
\end{figure}

In order to illustrate the evolution of the radiation entropy in the semiclassical model, let us consider two examples for the eigenvalue distribution of $\rho_{SC}$, both having the property that the participation ratio can be much smaller than the dimensionality of the matrix. First, we consider the simplest case when $\rho_{SC}$ has exactly $PR$ non-vanishing eigenvalues of magnitude $\overline\lambda$ and the remaining $(N - PR)$ eigenvalues are exactly vanishing. The corresponding distribution function reads
\be\label{dist_equal}
P(\lambda) = \frac{N - PR}N \, \delta(\lambda) + \frac{PR}N \, \delta\left(\lambda - \frac{N}{PR} \right) \;,
\ee
where the participation ratio is that of Eq.~\eqref{prsc}. The R\'enyi entropy in such a case is given by
\begin{multline}
H_2(N) = PR \, \ln{(1 + 2 \overline\lambda)} = \frac{N}{1 + N \cbh} \ln{(3 + 2 N \cbh)}\\
= S_{BH}(0) \frac{N}{S_{BH}(0)} \left( 1 - \frac{N}{S_{BH}(0)}\right) \ln{\left( 1+ \frac{2}{1-N/S_{BH}(0)} \right)} \;.
\end{multline}
The evolution of this radiation entropy with the number of emitted particles is presented in Figure~\ref{fig:MP}, which shows the predicted decline after the Page time.

Another eigenvalue distribution that illustrates the properties of $\rho_{SC}$ is the MP distribution~\eqref{MP}, as already discussed in the context of the Page model. The crucial difference from the Page model itself is that we now use the  MP distribution to model the single-particle density matrix $\rho_{SC}$ and not the reduced density matrix which represents the state of the emitted radiation. Requiring the participation ratio of the MP-distributed matrix~\eqref{MPPR} to be equal to the semiclassical prediction~\eqref{prsc}, one finds that the parameter $c$ of the MP distribution is given by
\be
c = \frac 1 {N \cbh} \; .
\ee
The R\'enyi entropy of the radiated particles can then be obtained by plugging the MP distribution~\eqref{MP} into Eq.~\eqref{renyi_dist}. The evolution of the entropy is shown in Figure~\ref{fig:MP}. One can observe that the entropy
is decreasing  after the Page time, just as  expected for the semiclassical model.

Although  these two examples were used for illustrative purposes, we do expect
the eigenvalue distribution for $\rho_{SC}$ to be similar in some aspects to the MP distribution. The moments of the distribution function of the eigenvalues of $\rho_{SC}$ are calculated and compared to the moments of the MP distribution in Appendix~\ref{sec:moments}. In spite of differing in details, the two sets of moments follow a similar pattern.

Having  established that the R\'enyi entropy of the  radiation in the semiclassical model starts to decrease after the Page time, let us speculate on a viable form for its functional dependence on the number of emitted particles. What  is known about the semiclassical model is that  $H_2(\rho_{SC})$   (1)
must  tend to the Hawking  value of $N\ln{3}$, as in Eq.~(\ref{H2H}), at early times and (2) must be symmetric under an exchange between the radiation and the BH, since the two reduced density matrices necessarily share the same set of eigenvalues. In particular, all physical quantities are required to be symmetric under the exchange of $N\leftrightarrow S_{BH}(N)$, as long as these quantities characterize the sizes of their respective Hilbert spaces $\mathcal H_{\textrm{out}}$ and $\mathcal H_{\textrm{BH}}$.

The participation ratio \eqref{prsc} is manifestly symmetric in just this way,
\be
PR(\rho_{SC})\;=\;\frac{N}{1+N\cbh(N)}\;=\;\frac{NS_{BH}(N)}{S_{BH}(N)+N} \;.
\ee
It is also of the same form as the participation ratio of the MP distribution \eqref{MPPR}, given that one identifies the parameters $n$ and $m$ of the Page model to be $N$ and $S_{BH}(N)$, respectively.\footnote{Note that, in the Page model, $n$ and $m$ are the dimensions of the reduced density matrices $\hat\rho_{out}$ and $\hat\rho_{P}$, and  are exponentially large numbers, $n \sim e^N$ and $m \sim e^{S_{BH}(N)}$.} This may seem peculiar because,  as discussed in Appendix~\ref{sec:moments}, the actual eigenvalue distribution of $\rho_{SC}$  differs from
the MP law. But, as far as we can tell, $\frac{NS_{BH}(N)}{S_{BH}(N)+N}$ is the simplest non-trivial, symmetric function which tends to $N$ for $N \ll S_{BH}(0)$, suggesting that such an identification is valid.

\begin{figure} [h]
\centering
\includegraphics[width = .7\linewidth]{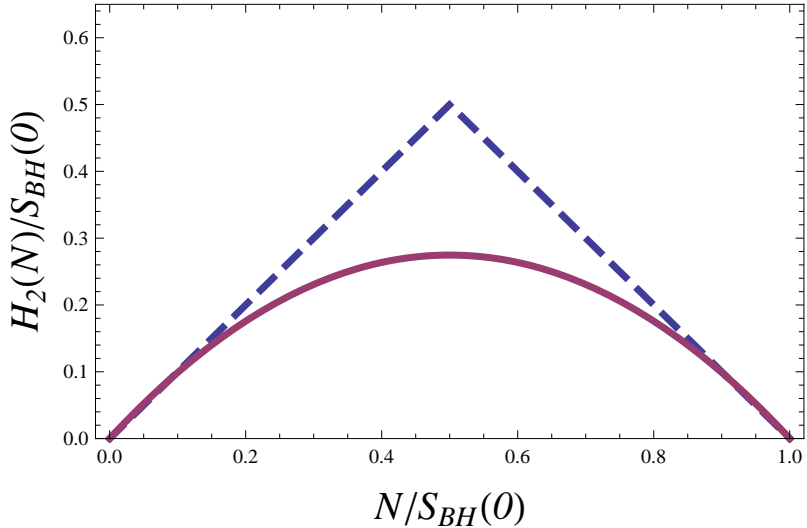}
\caption{Dependence of the R\'enyi entropy $H_2$ of the BH radiation on the number of emitted particles $N$. The solid purple line depicts the prediction of the semiclassical model from Eq.~\eqref{SCRenyi}  and the dashed blue line depicts the prediction of the Page model from Eq.~\eqref{PageRenyi}.\label{fig:Page}
}
\end{figure}

Motivated by these observations, our expectation is that the semiclassical R\'enyi entropy can be expressed in terms of the participation ratio as follows:
\be\label{SCRenyi}
H_2(\rho_{SC})\;=\; PR(\rho_{SC})\left[\ln{3}+c_{-1}\frac{1}{PR(\rho_{SC})}+ c_{-2}\frac{1}{(PR(\rho_{SC}))^2}+\cdots\right] \;,
\ee
where the coefficients are dimensionless numbers that are determined by the higher moments of the eigenvalue distribution. To see that  $c_0=\ln{3}$ is indeed the correct leading-order coefficient, consider that, at early times, $N \simeq PR$ to leading order in a $1/PR$ expansion. Thus, the average eigenvalue
$\overline{\lambda} = \frac{N}{PR} \simeq 1$ and the logarithm in $H_2$
then expands as $\ln{(1+2\overline{\lambda})} \simeq \ln{3}$.

The dependence~\eqref{SCRenyi} of the R\'enyi entropy on the number of
emitted  particles for  a semiclassical BH is presented in
Figure~\ref{fig:Page}.
Because the R\'enyi entropy depends only on  the participation
ratio $\frac{N}{1+N\cbh}$, one can determine the evolution of the former from that of the latter.
The participation ratio steadily grows until reaching a maximum at the Page time and then steadily
declines for the remainder of the evaporation process. This
is qualitatively similar behavior to the R\'enyi entropy for the Page model although, in the semiclassical model, the rate of purification---which is the rate of deviation from the linearly growing result of the Hawking's model---is actually faster than the Page-model rate~\cite{endgame}.

\section{Conclusion}\label{sec:conclusion}

We have exposed the relationship between the density matrix of the outgoing radiation from a BH and the corresponding single-particle density matrix in the case that the density matrix is approximately diagonal in mode-occupation number. It was then shown that the presence of the horizon leads to a high suppression of the off-diagonal elements in mode-occupation number of the density matrix for the emitted radiation. We have therefore concluded that the density matrix is approximately diagonal in mode-occupation number at all times. We have also  shown  how to regularize the infinities which arise as a consequence of Hawking's idealized picture of an eternal BH. It was   explained how the state of the emitted radiation from a BH is different from that of a standard black body of the same temperature. This analysis was then applied to three models of BH evaporation as a means of clarifying their differences  and contrasting their main features. Let us briefly summarize the main observations:\\

{\bf The Hawking model:}\\

The off-diagonal elements of the single-particle density matrix are highly suppressed for different frequencies and emission times, and those of the full density matrix are exponentially suppressed for differences in occupation numbers. The R\'enyi entropy \eqref{renyiH} scales with the number of emitted particles $N$ at all times, and so there is no possibility for the radiation
to be purified.\\

{\bf The Page Model:} \\

For a randomly chosen basis whose relationship to Hawking's basis is unspecified, the off-diagonal elements of the density matrix are uniform in magnitude with random phases. There is no hierarchy between different elements of the matrix. The R\'enyi entropy of the radiation \eqref{PageRenyi} increases linearly as $N$ until the Page time and decreases linearly as $S_{BH}(0)-N$ afterwards. Hence, purification is inevitable.\\

{\bf The semiclassical model:} \\

The off-diagonal elements of the single-particle density matrix are suppressed by a power of $S_{BH}^{-1/2}$ in a uniform way, whereas those of the density matrix are exponentially suppressed for different mode-occupation numbers in a similar way to the Hawking model. At the Page time, the contribution from the off-diagonal elements with regard to  frequency and emission time becomes significant because of
a  perturbative parameter that grows monotonically  with time.
The R\'enyi entropy \eqref{SCRenyi} scales with the participation ratio of the single-particle density matrix, which increases until the Page time and decreases thereafter. This suggests that the radiation starts to purify and does so at a rate which  is faster than that of the Page model. The question of how unitarity is recovered for a process of gravitational collapse  in the semiclassical model is discussed in detail in a companion paper~\cite{endgame}.

\section*{Acknowledgments}
The research of LA, RB, AK was supported by the Israel Science Foundation grant no. 239/10. The work of AK was also supported by the Kreitman foundation. The work of LA was also supported by the Minerva Foundation. The research of AJMM received support from an NRF Incentive Funding Grant 85353, an NRF Competitive
Programme Grant 93595 and Rhodes Research Discretionary Grants. AJMM thanks Ben-Gurion University for their hospitality during his visit.


\appendix

\section{Expression for the full density matrix in terms of $\rho^i_{\;j}$}\label{sec:app_formula}

In order to find an expression for the density matrix, it is useful to calculate a vacuum expectation value; namely,
\be\label{genfun}
\bra{0_-} e^{b^{+}_i \mu^i} e^{\lambda_j b^j} \ket{0_-} \;.
\ee
The resulting function of $\mu^i$ and $\lambda_j$ can be used as a generating function for the expectation values of any normal-ordered powers of the creation and annihilation operators $b^+_i$ and $b^j$.
The expectation value~\eqref{genfun} can be calculated by using \eqref{bogo} and applying the Baker--Campbell--Hausdorf (BCH) formula for the exponents:
\begin{multline}\label{gen_fun_1}
\bra{0_-} e^{b^{+}_i \mu^i} e^{\lambda_j b^j} \ket{0_-} 
=\bra{0_-} e^{ \mu^i\alpha_i^{\;j} a^+_j } e^{ - \mu^i\beta_{ij} a^j} e^{- \lambda_j\bar\beta^{jk} a^+_k} e^{ \lambda_j \bar\alpha^j_{\;k} a^k} \ket{0_-}\ e^{ -\half \mu^i\alpha_i^{\;j} \beta_{kj} \mu^k} e^{ -\half \lambda_i \bar\alpha^i_{\;k} \bar\beta^{jk}\lambda_j} \;.
\end{multline}

The exponential factors on the  outside of  the average arise from the commutators. By using Hawking's expressions in~\cite{Hawking:1976ra}
for the Bogolyubov coefficients $\alpha_{ij}$ and $\beta_{ij}$, one can show that \begin{equation}\label{ab}
\alpha_i^{\;j} \beta_{kj} = 0 \;, \qquad \bar\alpha^i_{\;k} \bar\beta^{jk} = 0\;,
\end{equation}
and so the external exponentials are equal to unity as their exponents
are vanishing.

Relations \eqref{ab} also ensure that only the matrix elements of $\widehat\rho^{\textrm{\; out}}$ between states with the same total occupation number are non-vanishing. But, as discussed in Section~\ref{sec:off_diagonal}, these relations hold only approximately for realistic BHs.

Returning to Eq.~\eqref{gen_fun_1}, one can see that the two outer factors within the expectation value disappear after acting on the vacuum $\ket{0_-}$. The remaining expression can be further simplified by applying the BCH relation once again:
\be
\bra{0_-} e^{ - \mu^i\beta_{ij} a^j} e^{- \lambda_j\bar\beta^{jk} a^+_k} \ket{0_-} = e^{\lambda_i \bar\beta^{ik}\beta_{jk}\mu^j} \cdot \bra{0_-} e^{- \lambda_j\bar\beta^{jk} a^+_k} e^{ - \mu^i\beta_{ij} a^j} \ket{0_-} = e^{\lambda_i \bar\beta^{ik}\beta_{jk}\mu^j} \;.
\ee
Finally, the answer for the generating function~\eqref{genfun} reads
\be\label{gf}
\bra{0_-} e^{b^{+}_i \mu^i} e^{\lambda_j b^j} \ket{0_-} = e^{\lambda_i \rho^i_{\;j}\mu^j} \;,
\ee
where $\rho^i_j$ is the single-particle density matrix as defined in Eq.~\eqref{rho}. For Hawking, $\rho^i_{\;j}$ is a diagonal matrix but, in more general setups, it can also have non-zero off-diagonal elements. In any case, $\rho^i_{\;j}$ is Hermitian and can thus be diagonalized by a unitary transformation.

We next look for a closed expression for the density matrix $\widehat \rho^{\textrm{out}}$ in terms of the $b$'s. In analogy with the harmonic oscillator, let us try the {\em ansatz}
\be\label{denmat_ansatz}
\widehat\rho^{\textrm{\; out}} = \frac1Z e^{-b^+_i \Omega^i_{\;j} b^j} \;,
\ee
where $\Omega$ is some $c$-number Hermitian matrix and $Z$ is the normalization factor.
The matrix $\Omega$ can be diagonalized by a unitary transformation, $\Omega = U^+ D U$, with $D$ being a real diagonal matrix. This allows us to introduce new annihilation operators $d^j \equiv U^j_{\ i} b^i$ for which the quadratic form $b^+\Omega\, b = b^+ U^+ D\, U \,b = d^+ D\, d$ is diagonal. Such a transformation is canonical, so that $\{d_i^+,d_i\}$ are also a good set of creation and annihilation operators.

Moreover, since this transformation does not mix the creation and annihilation operators between each other, the subspaces with a fixed total occupancy of $b$- and $d$-particles coincide, and the traces over the Fock spaces of $b$- and $d$-particles are equal. In particular, one can find the normalization $Z$ by summing over the states with a fixed occupation number of $d$-particles $\ket{n_i}$:
\begin{multline}\label{norm}
Z = \tr \left[ e^{-b^+_i \Omega^i_{\;j} b^j} \right] = \tr \left[ e^{-\sum_i d^+_i D^i_{\;i} d^i} \right] = \prod_i \sum_{n_i = 0}^\infty \bra{n_i} e^{-\sum_i D^i_{\;i} d^+_i d^i} \ket{n_i} = \\
= \prod_i \frac1{1 - e^{-D^i_{\;i}}} = \det \left[ \frac1{1 - e^{-D}}\right] = \det \left[ \frac1{1 - e^{-\Omega}}\right] \;.
\end{multline}
The last equality is due to the invariance of the determinant under unitary transformations of the matrix.

According to {\em ansatz}~(\ref{denmat}), the corresponding generating function \eqref{genfun} is
\be\label{genfunrho}
\la e^{b^{+} \mu} e^{\lambda\, b} \ra_{\widehat\rho} \equiv \frac1Z \tr{\left[ e^{b^{+} \mu} e^{\lambda\, b} e^{-b^+ \Omega\, b} \right]} \;.
\ee
If it is possible to choose $\Omega$ so that the result matches~\eqref{gf}, then ansatz~\eqref{denmat_ansatz} gives the correct expression for the density matrix.

 As before, the generating function~\eqref{genfunrho} can be calculated by tracing in the $d$-particle basis:
\begin{multline}
\frac1Z \tr{\left[ e^{b^{+} \mu} e^{\lambda\, b} e^{-b^+ \Omega\, b} \right]} = \frac1Z \tr{\left[ e^{d^{+}U\, \mu} e^{\lambda\,U^+ d} e^{-d^+ D\, d} \right]} = \\
= \prod_i \frac1{Z_i} \sum_{n_i = 0}^\infty \bra{n_i} {e^{d^{+}_i \l U\, \mu \r^i} e^{\l \lambda\,U^+\r_i\, d^i} e^{-d^+_i D^i_{\;i}\, d^i} }\ket{n_i} \;,
\end{multline}
where each of the factors in the bottom line corresponds to a single oscillator.

In order to evaluate the above expression, we will use the identity
\begin{equation}
\l 1 - e^{-D} \r \tr {\left[ e^{a^{+} \mu} e^{\lambda\, a} e^{-D\,a^+a} \right]} = \exp{ \l {\frac{\lambda \mu}{e^D - 1}} \r}\;,
\end{equation}
which is derived in~\cite{HOSC}. (For a web resource, see, {\em e.g.},~\cite{HOSC1}.) The expectation value~\eqref{genfunrho} then reads
\begin{multline}
\la e^{b^{+} \mu} e^{\lambda\, b} \ra_{\widehat\rho} = \prod_i \exp{ \l {\frac{\l \lambda\,U^+\r_i \l U\, \mu \r^i}{e^{D^i_{\;i}} - 1}} \r} = \exp{\l \sum_i {\frac{\l \lambda\,U^+\r_i \l U\, \mu \r^i}{e^{D^i_{\;i}} - 1}}\r} \equiv \\
\equiv \exp{\l \lambda\,U^+ \frac1{e^D - 1} \,U\, \mu \r } = \exp{\l \lambda\, \frac1{e^\Omega - 1} \,\mu \r } \;.
\end{multline}

Comparing the last expression to~\eqref{gf}, we find that the right-hand sides coincide if the matrix $\Omega$ is related to $\rho$ as
\be
\rho = \frac1{e^\Omega - 1} \;, \quad \text{or} \quad e^\Omega = 1 + \frac1\rho \;.
\ee

\section{Grey-body factors}\label{sec:app_greybody}

From the properties of the grey-body factors, one can see that the state of the field outside a large BH is quite different from the usual thermal state in an empty box, $\widehat\rho = \frac1 Z e^{-\beta \widehat H}$. The actual state of the radiation is that of a field in thermal equilibrium with the BH, which is very selective in what it absorbs and emits. For example, the occupation number of the modes with high angular momentum will be much less than what one would naively expect from just the Boltzmann suppression.

One way to present this modification away from thermality is to notice that, by definition, the grey-body factor $\Gamma_{\omega l m}$ is equal to the modulus square of the scattering amplitude of the corresponding mode on the BH. Therefore, it is physically intuitive to introduce a corresponding effective scattering cross-section (see, e.g., Ref.~\cite{Page:1976df}),
\be
\sigma_{\omega l m} = \frac\pi{\omega^2} \Gamma_{\omega l m} \;.
\ee

The rate of emission can then be written as
\be
\frac {N_{\omega l m}}{\delta t} = \int_0^\infty \frac{\omega^2 d \omega}{2 \pi^2} \frac{\sigma_{\omega l m}}{{e^{\omega/T}} -1} \;.
\ee
This is a three-dimensional thermal emission rate, where the geometric area of the emitter is different for each mode and is given by the absorption cross-section. In this language, the suppression of the high-$l$ modes comes from the fact that the effective area of the BH as seen by  them is much smaller than its geometric area.

Another way to see the difference from the usual thermal state is to notice that the density matrix of the equilibrium state has the form $\widehat \rho = Z^{-1}e^{-b^+ \Omega b}$ with $\Omega = \omega/T$ only if $\Gamma_{\omega l m}=1$. Indeed, the relation~\eqref{omegarho} between the single-particle density matrix $\rho=\frac{\Gamma_{\omega l m}}{e^{\omega/T}-1}$ and the matrix $\Omega$ implies that
\be
\Omega_{\omega l m} = \ln{\l \frac {e^{\omega/T} - (1 - \Gamma_{\omega l m})} {\Gamma_{\omega l m}}\r} \;.
\ee
Therefore, the density matrix for the radiation takes the form [cf. Eq.~\eqref{hawking}]
\be
\widehat \rho = \frac1Z \,\prod_{\omega l m} e^{-\Omega_{\omega l m} \widehat N_{\omega l m}} = \frac1Z \,\prod_{\omega l m}
\l \frac {\Gamma_{\omega l m}} {e^{\omega/T} - (1 - \Gamma_{\omega l m})} \r^{ \widehat N_{\omega l m}} \;.
\ee

For each given mode, the ratio of the probability of having $(N+1)$ particles to the probability of having $N$ particles is constant and given by
\be
\frac {P(N+1)} {P(N)} = \frac {\Gamma_{\omega l m}} {e^{\omega/T} - (1 - \Gamma_{\omega l m})} \;.
\ee
Meaning that, apart from the overall Boltzmann suppression factor, the emitted particle has a probability
of $(1 - \Gamma)$ to be scattered back into the BH,
\be
P(N+1) = e^{-\omega/T} [ \Gamma \, P(N) + (1- \Gamma) \, P(N+1) ] \;.
\ee

\section{Off-diagonal corrections to the Hawking model}\label{sec:offf_diagonal}

In order to estimate the off-diagonal corrections to the Hawking density matrix, we first need to consider the corrections to the products
\be\label{abc}
\bar\beta^{ik}\beta_{jk}\;, \qquad{\alpha_i}^j\beta_{kj}\;, \qquad{\bar\alpha^i}_k\bar\beta^{jk}\;,
\ee
where the second index is summed over. The first product defines the standard single-particle density matrix $\rho^i_j\equiv\bar\beta^{ik}\beta_{jk}$ and, in the case of Hawking's calculation, is diagonal and completely determines the density matrix of the outgoing radiation \eqref{denmat}. The products $\alpha\beta$ and $\bar\alpha\bar\beta$ arise in the expectation values of operators like
\begin{align}
&\bra{0_-}\left(b^+\right)^nb^{n+2k}\ket{0_-}\sim \left(\beta\bar\beta\right)^n\cdot\left(\alpha\beta\right)^{k}\;,\\
&\bra{0_-}\left(b^+\right)^{n+2k}b^{n}\ket{0_-}\sim \left(\beta\bar\beta\right)^n\cdot\left(\bar\alpha\bar\beta\right)^{k}\;
\end{align}
and, thus, define the elements of the density matrix between states that differ in total occupation
number, $\Delta {\cal N}=2k$.

The products $\alpha\beta$ and $\bar\alpha\bar\beta$ also appear during the computation of the generating function \eqref{gen_fun_1}. Consequently, the sole dependence of the density matrix on
the single-particle density matrix $\rho^i_j$ is only truly valid in the case when the products $\alpha\beta$ and $\bar\alpha\bar\beta$ vanish. This assumption is true for Hawking's idealized calculation but, as shown below,
 modified for physically realistic BHs.

In order to find the corrections to the products \eqref{abc}, we rewrite the definition of the coefficients $\alpha_{\omega\omega'}$ and $\beta_{\omega\omega'}$ in the basis of Fourier modes,
\begin{align}
&\alpha_{\omega\tilde\omega}=if_\alpha(\omega)\left(\tilde\omega\right)^{-1/2+\frac{i\omega}{\kappa}}e^{i(\omega-\tilde\omega)v_0}\;,\\
&\beta_{\omega\tilde\omega}=-if_\beta(\omega)\left(\tilde\omega\right)^{-1/2+\frac{i\omega}{\kappa}}e^{i(\omega+\tilde\omega)v_0}\;,
\end{align}
where
\begin{align}
f_\alpha(\omega)\equiv\frac{t_\omega}{2\pi}\frac{1}{\sqrt{\omega}}\,\Gamma\left(1-\frac{i\omega}{\kappa}\right)e^{\frac{\pi\omega}{2\kappa}}\;,\qquad f_\beta(\omega)\equiv f_\alpha(\omega)e^{-\frac{\pi\omega}{\kappa}}\;,
\end{align}
and $v_0$ denotes the position of the BH horizon in advanced time, $\kappa=2\pi T$ is the BH surface gravity and $t_\omega$ is the transmission coefficient for which $|t_\omega|^2=\Gamma_\omega$.

The products \eqref{abc} can be rewritten in the wave-packet basis of Section~\ref{sec:state}. For instance,
\begin{align}
{\left(\bar\beta\beta\right)^{jn}}_{j'n'}=\eps^{-1} \int_{j \eps}^{(j+1) \eps} d \omega \, e^{2\pi i \, n \, \omega/\eps} \int_{j' \eps}^{(j'+1) \eps} d \omega' \, e^{- 2\pi i \, n' \, \omega'/\eps} \rho_{\bar\beta\beta}(\omega,\omega')\;,
\end{align}
where
\be \label{rho-beta-beta}
\rho_{\bar\beta\beta}(\omega,\omega')\equiv\int_0^\infty d\tilde\omega\,\bar\beta_{\omega\tilde\omega}\beta_{\omega'\tilde\omega}\;.
\ee
There are similar expressions for ${\left(\alpha\beta\right)}_{jn\,j'n'}$ and ${\left(\bar\alpha\bar\beta\right)^{jn\,j'n'}}$. The product (\ref{rho-beta-beta})
and its analogues
can  be expressed  in terms of the Fourier-mode basis as
\begin{align}
&\rho_{\bar\beta\beta}(\omega,\omega')=\bar{f_\beta}(\omega)f_\beta(\omega')e^{i(\omega'-\omega)v_0}\int_{-\infty}^{+\infty} dy \,e^{iy/\kappa(\omega'-\omega)}\;,\\
&\rho_{\alpha\beta}(\omega,\omega')=f_\alpha(\omega)f_\beta(\omega')e^{i(\omega'+\omega)v_0}\int_{-\infty}^{+\infty} dy \,e^{iy/\kappa(\omega'+\omega)}\;,\\
&\rho_{\bar\alpha\bar\beta}(\omega,\omega')=\bar{f_\alpha}(\omega)\bar{f_\beta}(\omega')e^{-i(\omega'+\omega)v_0}\int_{-\infty}^{+\infty} dy \,e^{-iy/\kappa(\omega'+\omega)}\;,
\end{align}
where the integration variable has been changed to $y=\ln(\tilde\omega)$.

Substituting the previous set of relations into the corresponding expressions for the products in the wave-packet basis  and then integrating over the frequencies $\omega$ and $\omega'$, we arrive at
\begin{align}
&{\left(\bar\beta\beta\right)^{jn}}_{j'n'}=f_\beta\left(\bar\omega'\right)\bar{f_\beta}(\bar\omega)\frac{2\kappa}{\pi}\int_{-\infty}^{+\infty} d\mu\, \frac{e^{2\pi i\mu(j'-j)}}{(\mu-\tilde n)(\mu-\tilde n')}\sin^2(\pi\mu)\;,\\
&{\left(\alpha\beta\right)}_{jn\,j'n'}=f_\alpha\left(\bar\omega'\right){f_\beta}(\bar\omega)\frac{2\kappa}{\pi}\int_{-\infty}^{+\infty} d\mu\, \frac{e^{2\pi i\mu(j'+j+1)}}{(\mu-\tilde n)(\mu-\tilde n')}\sin^2(\pi\mu)\;,\\
&{\left(\bar\alpha\bar\beta\right)^{jn\,j'n'}}=\bar{f_\alpha}\left(\bar\omega'\right)\bar{f_\beta}(\bar\omega)\frac{2\kappa}{\pi}\int_{-\infty}^{+\infty} d\mu\, \frac{e^{-2\pi i\mu(j'+j+1)}}{(\mu-\tilde n)(\mu-\tilde n')}\sin^2(\pi\mu)\;,
\end{align}
where $\bar\omega\equiv\eps\left(j+1/2\right)$ and $\bar\omega'\equiv\eps\left(j'+1/2\right)$ are the mean values of the frequencies in the given range of integration, $\tilde n\equiv n-n_0$, $\tilde n'\equiv n'-n_0$ with
$n_0$ defined by $v_0=2\pi n_0\eps^{-1}$, and the integration variable has been changed to $\mu=y\eps/(2\pi\kappa)$. We see that the three integrals differ only by the power in the exponent, and so it is convenient to  define
\be\label{js}
J_{\bar\beta\beta}\equiv j'-j\;,\qquad J_{\alpha\beta}\equiv j'+j+1\;,\qquad J_{\bar\alpha\bar\beta}\equiv -(j'+j+1)\;.
\ee

Let us now introduce infrared and ultraviolet cutoffs for the integration range of the frequencies, as discussed in
the Subsection~\ref{sec:off_Hawking},
\be
\int_{-\infty}^{+\infty}d\mu\quad\to\quad\int _{-\mu_*}^{\mu_*}du\;.
\ee
Due to the similarity of the three integrals above, we need only to consider one,
\be\label{integral}
I_{\Delta n}(J)=\int_{-\mu_*}^{+\mu_*} d\mu\, \frac{e^{2\pi i\mu J}}{(\mu-\tilde n)(\mu-\tilde n')}\sin^2(\pi\mu)\;,
\ee
and then interpret the results for the different choices of $J$ in Eq.~\eqref{js}.
The integral does need, however, to be treated differently for the cases with $\Delta n\equiv n'-n=0$ and $\Delta n\neq 0$,
as well as for the cases with $J=0$ and $J\neq 0$.

\subsection {Off-diagonal elements in frequency}
When $\Delta n=0$, the integral~(\ref{integral}) becomes
\be
I_0(J)=\int_{-\mu_*}^{\mu_*}d\mu\,\frac{e^{2\pi i\mu J}}{(\mu-\tilde n)^2}\sin^2(\pi\mu)\;.
\ee

For the case $J=0$, the previous integral can be evaluated to give
\be
I_0(0)=-\frac{\sin^2(\pi\mu_*)}{\mu_*-\tilde n}-\frac{\sin^2(\pi\mu_*)}{\mu_*+\tilde n}+\pi\textrm{Si}(2\pi(\mu_*-\tilde n))+\pi\textrm{Si}(2\pi(\mu_*-\tilde n))\;,
\ee
where $\textrm{Si}(x)=\int_0^xdt\,\sin t/t$ is the sine integral. It has the
large-$x$ expansion
\be
\textrm{Si}(x)=\frac{\pi}{2}-\frac{\cos x}{x}-\frac{\sin x}{x^2}+\mathcal O(x^{-3})\;.
\ee
Hence, in the limit when $\mu_*\gg \tilde n$, the integral becomes
\be\label{int0}
I_0(0)=\pi^2-\frac{1}{\mu_*}-\frac{\sin(2\pi\mu_*)}{2\pi\mu_*^2}+\mathcal O(\mu_*^{-3})\;.
\ee

When $J\neq 0$, the integral can be split as
\be\label{int_split}
I_0(J)=\frac{1}{2}\left(h(J)-\frac{1}{2}h(J-1)-\frac{1}{2}h(J+1)\right)\;,
\ee
where
\be
h(J)=\int_{-\mu_*}^{\mu_*}d\mu\,\frac{e^{2\pi i \mu J}}{(\mu-\tilde n)^2}\;.
\ee

We find that the last integral is expressible as
\be
h(J)=-\frac{e^{2\pi i\mu_*J}}{\mu_*-\tilde n}-\frac{e^{-2\pi i \mu_*J}}{\mu_*+\tilde n}-2\pi i J\textrm{E}_1(-2\pi iJ(\mu_*-\tilde n))+2\pi iJ\textrm{E}_1(2\pi i J(\mu_*+\tilde n))\;,
\ee
where $E_1(z)=\int_1^\infty dt\,e^{-zt}/t$ is the exponential integral with the large-$x$ expansion
\be
\textrm E_1(ix)=e^{-ix}\left(\frac{1}{ix}+\frac{1}{x^2}+\mathcal O(x^{-3})\right)\;.
\ee
In the limit when $\mu_*\gg \tilde n$, the function $h(J)$ then becomes
\be\label{res1}
h(J)=\left\{
\begin{matrix}
-\frac{2}{\mu_*}+\mathcal O(\mu_*^{-3}),\quad J=0\;,\\
\\
\frac{2}{\mu_*}\frac{\sin(2\pi\mu_*J)}{2\pi\mu_* J}+\mathcal O(\mu_*^{-3})\,,\quad J\neq0\;.
\end{matrix}
\right.
\ee

The integral $I_0(J)$ at a given value of $J$ can then be evaluated by combining Eqs.~\eqref{int0},~\eqref{int_split} and~\eqref{res1}. For example, $I_0(1)$ works out to be
\begin{align}
I_0(1)&=\frac{1}{2}\left(h(1)-\frac{1}{2}h(0)-\frac{1}{2}h(2)\right)\nonumber\\
&=\frac{1}{2}\left(\frac{\sin(2\pi\mu_*)}{\pi\mu_*^2}+\frac{1}{\mu_*}-\frac{1}{4}\frac{\sin(4\pi\mu_*)}{\pi\mu_*^2}\right)\;.
\end{align}

\subsection{Off-diagonal elements in mode-occupation number}
When $\Delta n\neq 0$, the integral~(\ref{integral}) can be split as
\be\label{int_split2}
I_{\Delta n}(J)=\frac{1}{2}\left(g(J)-\frac{1}{2}g(J-1)-\frac{1}{2}g(J+1)\right)\;,
\ee
where
\be
g(J)=\frac{1}{\Delta n}\int_{-\mu_•}^{\mu_*}d\mu\left(\frac{e^{2\pi iJ}}{\mu-\tilde n'}-(\tilde n'\leftrightarrow\tilde n)\right)\;.
\ee

If $J=0$, the integral can be computed exactly. For $J\neq 0$, it is possible to redefine the integration variables in the two summands so as to move the dependence on the expansion parameter $\mu_*$ from the limits to the integrand:
\be
g(J)=\frac{2\cos(2\pi\mu_*J)}{\Delta n\mu_*}\int_{\tilde n'}^{\tilde n}dt\,\frac{e^{-2\pi it J}}{1-\left(\frac{t}{\mu_*}\right)^2}+\frac{2i\sin(2\pi\mu_*J)}{\Delta n\mu_*}\int_{\tilde n'}^{\tilde n}dt\,\frac{e^{-2\pi it J}}{1-\left(\frac{t}{\mu_*}\right)^2}\frac{t}{\mu_*}\;.
\ee
One can then expand the integrand in the limit when $\mu_*\gg \tilde n,\tilde n'$ and evaluate the integral to
obtain
\be\label{res2}
g(J)=\left\{
\begin{matrix}
-\frac{2}{\mu_*}+\mathcal O(\mu_*^{-3})\;,\quad J=0\;,\\
\\
\frac{2}{\mu_*}\frac{\sin(2\pi\mu_*J)}{2\pi\mu_* J}+\mathcal O(\mu_*^{-3})\;,\quad J\neq0\;.
\end{matrix}
\right.
\ee

Comparing $g(J)$ to the function $h(J)$ in Eq.~\eqref{res1}, we see that the two functions are equivalent, $g(J)=h(J)$. Hence, for $J\neq 0$, the two cases are coincident,
\be
I_{\Delta n}(J) =I_{0}(J)\;,\qquad J\neq 0\;.
\ee

On the other hand, at $J=0$, the integral $I_{\Delta n}$ differs from $I_0(0)$ in Eq.~\eqref{int0}
only by a zeroth-order, diagonal term:
\begin{align}
I_{\Delta n}(0)&=\frac{1}{2}\left(h(0)-\frac{1}{2}h(-1)-\frac{1}{2}h(1)\right)\nonumber\\
&=-\frac{1}{\mu_*}-\frac{\sin(2\pi\mu_*)}{2\pi\mu_*^2}+\mathcal O(\mu_*^{-3})
\nonumber \\
&= I_{0}(0)-\pi^2\;.
\end{align}

Hence, the corrections in both cases, $\Delta n=0$ and $\Delta n\neq 0$, coincide and are uniform in $\Delta n$.

\subsection{Final Result}
Up to the leading order in $\mu_*^{-1}$, the integral \eqref{integral} then becomes
\be
I_{\Delta n}(J)=\pi^2\delta_{\Delta n, 0}\,\delta _{J,0}+\frac{1}{\mu_*}\left(-\delta _{J,\,0}+\frac{1}{2}\delta_{J,\,1}+\frac{1}{2}\delta_{J,\,-1}\right)+\mathcal O(\mu_*^{-2})\;,
\ee
where the second-order corrections are subleading for $J=0,\pm1$. In the other cases---{\em i.e.}, when $J\neq 0,\pm1$--- it is the corrections at  second order in $\mu_*^{-1}$  that are  dominant:
\be
I_{\Delta n}(J)=\frac{1}{\mu_*}\left(\frac{\sin(2\pi\mu_*J)}{2\pi\mu_*J}-\frac{1}{2}\frac{\sin(2\pi\mu_*(J-1))}{2\pi\mu_*(J-1)}-\frac{1}{2}\frac{\sin(2\pi\mu_*(J+1))}{2\pi\mu_*(J+1)}\right)+\mathcal O(\mu_*^{-3})\;.
\ee

In order to interpret these results as corrections to the density-matrix elements, one has to employ the different definitions of $J$ in Eq.~\eqref{js} for the different cases.

 \subsubsection*{Interpretation}
The matrix elements of the product ${\left(\bar\beta\beta\right)^{jn}}_{j'n'}$
are  proportional to
\be
{\left(\bar\beta\beta\right)^{jn}}_{j'n'}\sim I_{\Delta n}(J_{\bar\beta\beta})\;,\qquad J_{\bar\beta\beta}=j'-j\;.
\ee
Since the corrections are found to be uniform in $\Delta n=n'-n$,  we
can  represent the matrix ${\left(\bar\beta\beta\right)^{jn}}_{j'n'}$ as a tensor product of $j$'s and $n$'s. Then  each matrix element that is labeled
by  $j$ and $j'$ is itself a uniform matrix in terms of  $n$ and $n'$. In this way, the product can be represented in the following form:
\be
{\left(\bar\beta\beta\right)^{jn}}_{j'n'}\sim\begin{pmatrix}
1&\frac{1}{\mu_*}&&&\\
\frac{1}{\mu_*}&1&\frac{1}{\mu_*}&&{\mbox{\Large $\frac{1}{\mu_*^2J}$}}\\
&\ddots&\ddots&\ddots&\\
{\mbox{\Large $\frac{1}{\mu_*^2J}$}}&&\frac{1}{\mu_*}&1&\frac{1}{\mu_*}\\
&&&\frac{1}{\mu_*}&1
\end{pmatrix},
\ee
where each depicted entry represents  a uniform block and only the order of magnitude of the leading term is shown.

For the products $\alpha\beta$ and $\bar\alpha\bar\beta$, the corresponding index is $J_{\alpha\beta}=\pm(j'+j+1)$, respectively. Since both $j$ and $j'$ are positive, these products have no zeroth-order contributions and only the matrix elements with $j=j'=0$ have corrections of the order $\mathcal O(1/\mu_*)$. All the other entries of the matrices $\alpha\beta$ and $\bar\alpha\bar\beta$ receive only second-order corrections $\mathcal O\left(1/(\mu_*^2J)\right)$.

How these findings impact upon  the off-diagonal elements of the full density
matrix is discussed in Subsection~\ref{sec:off_Hawking} of the main text.

\section{Higher moments of the Marchenko--Pastur  and $\rho_{SC}$ distributions}\label{sec:moments}

Here, we will  quantify more precisely the differences between the
Marchenko--Pastur (MP) distribution and the eigenvalue distribution of $\rho_{SC}$. This entails assigning the
distributions  the same participation ratio and then determining  how their higher moments are different.

Comparing their respective participation ratios (with $n=N$) in Eqs.~\eqref{MPPR} and~\eqref{prsc}, one can readily identify the parameter $c$ for the semiclassical two-point
function,
\be\label{c}
c \;=\; \frac 1 {N \cbh(N)} \; .
\ee

Let us consider the case $N\cbh\gtrsim 1$. This is really the regime
of interest, since smaller values of $N\cbh$ correspond to the case in which the Hawking model is valid (up to small corrections). We will further assume that the phases of the off-diagonal terms in $\rho_{SC}$ can be treated as random.
One can then estimate
the higher moments of $\rho_{SC}$ up to combinatorial factors and sub-leading terms in small $1/N\cbh$, which leads to
\bea
\frac{\tr \left(\rho_{SC}^{2p}\right)}{(\tr \rho_{SC})^{2 p}} &\simeq& \frac{N^{p+1}\left( \cbh \right)^{p}}{N^{2 p}}= \frac{\left( \cbh \right)^{p}}{N^{ p-1}}\;,
\label{moments_even} \\
\frac{\tr\left(\rho_{SC}^{2p+1}\right)}{(\tr \rho_{SC})^{2 p+1}}&\simeq&
\frac{N^{p+1}\left( \cbh \right)^{p}\; N}{N^{2p+1}}=\frac{\left( \cbh \right)^{p}\; }{N^{p-1}}\;.
\label{moments_odd}
\eea

The basic idea behind these estimates is that the off-diagonal parts of the matrices are dominant when $N \cbh\gtrsim 1$ and the randomness of the phases requires these parts to sum coherently ({\em i.e.}, restricted to sums of the form $\sum_{ij} M_{ij}M_{ji}$). A simple example should suffice to illustrate the point. Let $\gamma$ and $\eta$ be the diagonal and off-diagonal parts respectively of a matrix $\rho$. Applying the rule of coherent summation and, otherwise, insisting on the maximum power of $\eta$, we have, for the $p=2$ case,
\bea
\tr\rho^4 &=& \sum_{ijkl}\rho_{ij}\;\rho_{jk}\;\rho_{kl}\;\rho_{li}\;\sim\; \sum_{ijl}\eta_{ij}\;\eta_{ji}\;\eta_{il}\;\eta_{li}\;, \\
\tr\rho^5 &=& \sum_{ijklm}\rho_{ij}\;\rho_{jk}\;\rho_{kl}\;\rho_{lm}\;\rho_{mi} \;\sim\;\sum_{ijl}\eta_{ij}\;\eta_{ji}\;\eta_{il}\;\eta_{li}\;\gamma_{ii}\;.
\eea
As one can see, each of these traces results in $3=p+1$ independent summations,
a trend which continues for any value of $p$. This accounts for the
 factors of $N^{p+1}$ in Eqs.~(\ref{moments_even}) and~(\ref{moments_odd}); the rest
is determined by the magnitude of the elements.
These results indicate that the higher moments of the eigenvalue distribution are determined by an expansion in $\sqrt{\cbh}$.

The moments of the MP distribution, on the other hand, are expressed as a power series in $c$ and not square roots thereof. Indeed, for the same conventions and the same regime of small $c=1/(N \cbh)$,
the MP distribution would yield for the high moments
\be
\frac{\tr (\rho_{MP}^{n})}{(\tr \rho_{MP})^n} \;\simeq\; \left(\frac{1}{c}\right)^{n-1} \left(\frac{1}{N}\right)^{n-1} = \left( \cbh\right)^{n-1}\;,\;\;\;\;\;{\rm for}\;\;\;\;\;n\gg 1\;.
\ee
This follows from the observation that the MP distribution~\eqref{MP} has, for small values of $c$, about $N c\sim 1/\cbh$ large and (roughly) equal-valued eigenvalues $\lambda\sim 1/c$.

This discrepancy between the MP distribution and the semiclassical distribution is a consequence of the square root of $\cbh$ appearing in the off-diagonal elements of $\rho_{SC}$. Hence, $\rho_{SC}$ does not precisely conform to an MP distribution nor should it necessarily be expected to. However, when $N \cbh \sim 1$, both expressions for the moments of the distributions scale in the same
way, $\frac{\tr (\rho^{n})}{(\tr \rho)^n}\sim (\cbh)^{n-1}$. Therefore, we do expect that the two distributions share the same general features; in particular, once $c=1$ ($N \cbh=1$) is reached, both eigenvalue distributions begin to develop support for zero eigenvalues.

Yet, when $N \cbh \gg 1$ is true, the higher moments of the eigenvalue distribution of $\rho_{SC}$ are much more suppressed than those of the $MP$-distribution.
For instance, the $n^{\rm th}$ semiclassical  moment is smaller by a relative factor $1/(N \cbh)^{n/2}$ than its MP counterpart. Nevertheless, we do expect
that, in this case, both distributions have only a few large eigenvalues but apparently differ in the detail. It would be interesting to find out what is the actual eigenvalue distribution of the semiclassical matrix.

\end{document}